\documentclass[%
 reprint,
 superscriptaddress,
 amsmath,amssymb,
pra,
]{revtex4-1}
\usepackage{graphicx}% Include figure files
\usepackage{dcolumn}% Align table columns on decimal point
\usepackage{bm}% bold math
\usepackage{braket} %para escribir bra-kets
\usepackage[caption=false]{subfig} % <=== "caption=false" added
\usepackage{changepage}
\usepackage{float}
\usepackage{amsmath}
\usepackage{xcolor}

\begin{document}

%\preprint{APS/123-QED}

\title{Resonant helicity mixing of electromagnetic waves propagating through matter}% Force line breaks with \\

\author{Jon Lasa-Alonso}
\altaffiliation[jonqnanolab@gmail.com]{}
\affiliation{Centro de F\'isica de Materiales, Paseo Manuel de Lardizabal 5, 20018 Donostia-San Sebasti\'an, Spain}
\affiliation{Donostia International Physics Center, Paseo Manuel de Lardizabal 4, 20018 Donostia-San Sebasti\'an, Spain}
\author{Jorge Olmos-Trigo}
\affiliation{Donostia International Physics Center, Paseo Manuel de Lardizabal 4, 20018 Donostia-San Sebasti\'an, Spain}
\author{Chiara Devescovi}
\affiliation{Donostia International Physics Center, Paseo Manuel de Lardizabal 4, 20018 Donostia-San Sebasti\'an, Spain}
\author{Pilar Hern\'andez}
\affiliation{IFIC (CSIC-UVEG), Edificio Institutos Investigaci\'on, Apt. 22085, 46071 Valencia, Spain}
\author{Aitzol Garc\'ia-Etxarri}
\affiliation{Donostia International Physics Center, Paseo Manuel de Lardizabal 4, 20018 Donostia-San Sebasti\'an, Spain}
\affiliation{IKERBASQUE, Basque Foundation for Science, Mar\'ia D\'iaz de Haro 3, 48013 Bilbao, Spain}
\author{Gabriel Molina-Terriza}
\altaffiliation[gabriel.molina.terriza@gmail.com]{}
\affiliation{Centro de F\'isica de Materiales, Paseo Manuel de Lardizabal 5, 20018 Donostia-San Sebasti\'an, Spain}
\affiliation{Donostia International Physics Center, Paseo Manuel de Lardizabal 4, 20018 Donostia-San Sebasti\'an, Spain}
\affiliation{IKERBASQUE, Basque Foundation for Science, Mar\'ia D\'iaz de Haro 3, 48013 Bilbao, Spain}

\date{\today}

\begin{abstract}
Dual scatterers preserve the helicity of an incident field, whereas antidual scatterers flip it completely. In this setting of linear electromagnetic scattering theory, we provide a completely general proof on the non-existence of passive antidual scatterers. However, we show that scatterers fulfilling the refractive index matching condition flip the helicity of the fields very efficiently without being in contradiction with the law of energy conservation. Moreover, we find that this condition is paired with the impedance matching condition in several contexts of electromagnetism and, in particular, within Fresnel's and Mie's scattering problems. Finally, we show that index-matched media induce a resonant helicity mixing on the propagating electromagnetic waves. We reach to this conclusion by identifying that the refractive index matching condition leads to the phenomenon of avoided level-crossing. Our contribution not only closes a historical discussion within the Nanophotonics community, but also opens up new possibilities in the fields of Metamaterials and Particle Physics.
\end{abstract}

\maketitle

%\tableofcontents

\section{Introduction}

The coupling of two harmonic oscillators is one of the most studied systems in Physics textbooks with long-stretching implications in many areas of research \cite{Allen_Eberly}. Under suitable modifications, coupled harmonic oscillators allow us to understand such distant phenomena as the strong-coupling of light matter interactions \cite{Novotny_StrongCoupling}, hybridization of plasmonic and polaritonic systems \cite{StrongCoupling1, StrongCoupling2}, atomic dark states \cite{Tripod1, Tripod2} or even neutrino oscillations \cite{MSW0, MSW1, MSW2}. Here we explore a hitherto hidden form of this kind of coupling which appears within the polarization of electromagnetic waves propagating through matter.

\begin{figure}[b]
    \centering
    \includegraphics[width = 0.40\textwidth]{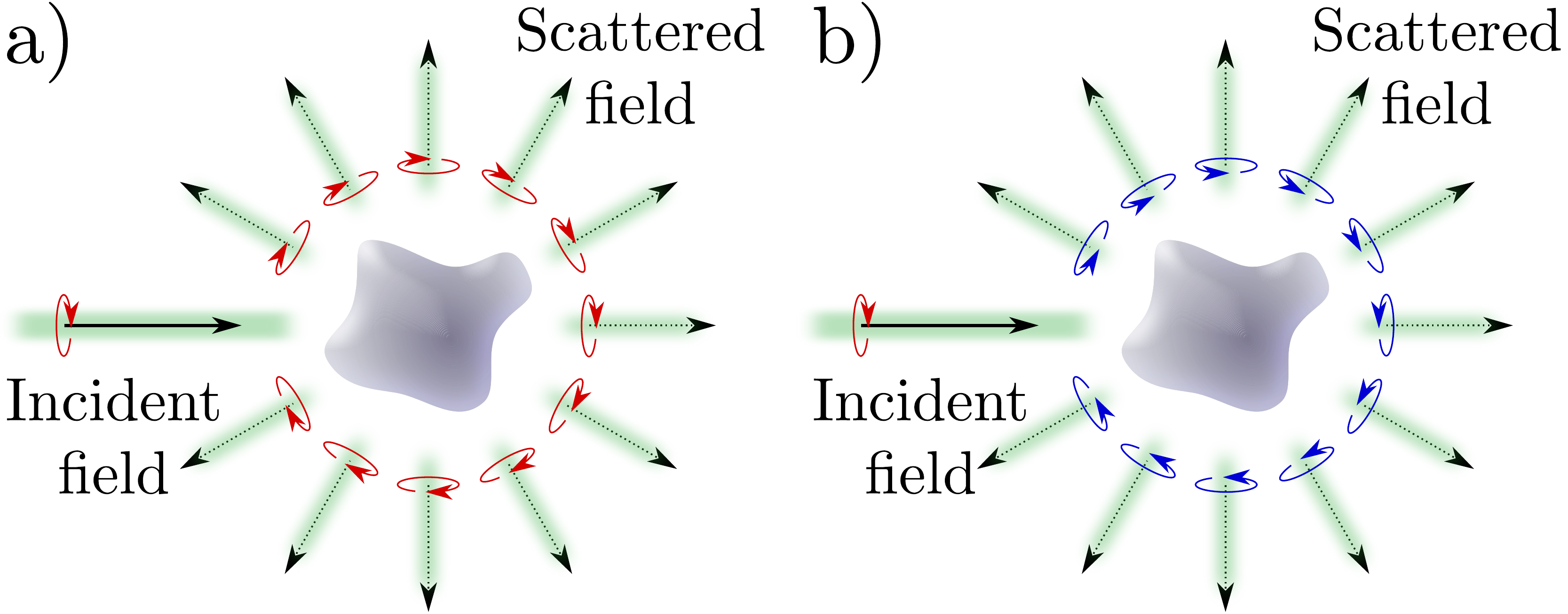}
    \caption{a) Sketch of a dual sample, i.e. one which preserves the helicity of an incident field. b) Sketch of an antidual sample, i.e. one which completely flips the helicity of an incident field.}
    \label{DualAntidual}
\end{figure}

A natural way of describing the polarization of electromagnetic fields is to use helicity as a degree of freedom. The electromagnetic helicity, $\Lambda$, can be understood as the circular polarization basis in the plane wave expansion of an electromagnetic field. In an operator formalism, helicity is defined as the projection of the spin angular momentum, $\mathbf{S}$, onto the direction of the linear momentum, $\mathbf{P}$, i.e. $\Lambda = \mathbf{S}\cdot\mathbf{P}/|\mathbf{P}|$, and for monochromatic electromagnetic fields in homogeneous media it takes the particularly simple form of $\Lambda = k^{-1}\nabla \times$ \cite{Birula3, Birula1, Messiah, PRLMolina, BliokhHel}, where $k$ is the modulus of the wavevector. The electromagnetic helicity is conserved in the interaction with dual samples, i.e. systems that have the same response to electric and magnetic fields. Thus, a dual sample, when illuminated with a beam with well-defined helicity, produces a scattered electromagnetic field with the same helicity of the incident field (see Fig. \ref{DualAntidual}a). On the other hand, an antidual sample produces a scattered electromagnetic field only with the opposite helicity (see Fig. \ref{DualAntidual}b). These concepts have been essential to understand the anomalous scattering of light by spherical particles \cite{KerkerRef01, KerkerRef1} and nanodisks \cite{KerkerRef2}. Such scattering phenomena were first described by Kerker and coworkers in what they are now known as the Kerker conditions \cite{Kerker}. It has been shown that the absence of backscattered field, in the first Kerker condition, is directly related to dual spherical scatterers. On the other hand, the second Kerker condition, which leads to the zero forward scattering condition, has been shown to be related to antidual particles \cite{ZambranaKerker}. While the properties of dual particles are well understood and have been exploited for a wide variety of applications both experimentally and theoretically \cite{KerkerApp0, KerkerApp1, KerkerApp2, KerkerApp3, ACSSolomon, ACSLasa, Bliokh1, Barnett1, SymProt}, antidual particles remain elusive. Even if similar scientific efforts have been dedicated, difficulties have arisen when trying to identify experimentally realizable antidual scatterers \cite{Optimal3, Optimal31, Optimal32, CamaraExcep, AluEngheta, Optimal4, KerkerRef0, MiroZero,  DualAntidualModes, ShenGain, CrisAnapole, RAliGain}.

In this work, we demonstrate in full generality that antidual scatterers which are not externally pumped with energy are precluded in linear electromagnetic scattering theory. However, we show that modes of electromagnetic helicity can be maximally mixed under a resonant condition that respects the law of energy conservation. Such a resonant condition appears in a symmetric manner with respect to the duality condition, i.e. just as a dual medium occurs when $\nabla Z = 0$, where $Z$ is the local impedance, the resonant helicity mixing condition is found for media in which $\nabla n = 0$, where $n$ is the local refractive index.  This closes a long-standing controversy about the implementation of helicity flipping scatterers, while at the same time, we indicate where the experimental efforts should be directed instead. Finally, we characterize the resonant helicity mixing condition with the aid of time-independent perturbation theory and find that it can be related to the phenomenon of avoided level-crossing. The connection with quantum two-level systems aligns the resonant helicity mixing effect with other relevant physical phenomena in Particle Physics. 

\section{Antiduality and energy conservation} \label{Sec2}
Let us first show that passive antidual scatterers are not compatible with energy conservation in linear electromagnetic scattering theory. To that end, let us consider the most general situation in which an arbitrary incident field with a definite frequency $\omega$, $[\mathbf{E}_i(\mathbf{r}, \omega), \mathbf{H}_i(\mathbf{r}, \omega)]$, scatters off a sample which, in turn, emits a field at the same frequency, $[\mathbf{E}_s(\mathbf{r}, \omega), \mathbf{H}_s(\mathbf{r}, \omega)]$. With such fields one naturally computes the total fields outside the scatterer just by taking the sum of the incident and the scattered electromagnetic fields, i.e. $\mathbf{E}_t = \mathbf{E}_i + \mathbf{E}_s$ and $\mathbf{H}_t = \mathbf{H}_i + \mathbf{H}_s$. With the total fields, the absorbed, scattered, or extincted powers are computed in terms of the time-averaged Poynting vector, $\mathbf{S}_t = \frac{1}{2}\text{Re}[\mathbf{E}_t\times\mathbf{H}_t^*]$ \cite{Bohren, Novotny, Jackson}. From Poynting's theorem, the absorbed power is calculated through the integral of the flux of the total Poynting vector across a surface $S$ which encloses the sample, i.e. $W_a = -\int \mathbf{S}_t\cdot \mathbf{n}~dS$, where $\mathbf{n}$ is a unitary vector perpendicular to the surface. By expanding $\mathbf{S}_t$ in terms of the incident and scattered fields, one obtains the most general form of the energy conservation law in electromagnetic scattering theory: $W_a = W_{ext} - W_s$, where $W_{ext}$ and $W_s$ are the extincted and scattered powers, respectively. The extincted and scattered powers also depend on the flux of a vector, in a similar fashion to $W_a$, but using $\mathbf{S}_{ext} = -\frac{1}{2}\text{Re}[\mathbf{E}_i \times \mathbf{H}_s^* + \mathbf{E}_s \times \mathbf{H}_i^*]$ for $W_{ext}$ and $\mathbf{S}_s = \frac{1}{2}\text{Re}[\mathbf{E}_s \times \mathbf{H}_s^*]$ for $W_s$.

Now, we will show that $\mathbf{S}_{ext}$ vanishes identically for a generic antidual sample. Indeed, an antidual scatterer is most generally defined as one which completely flips the helicity of the incident electromagnetic field. This implies that given an incident illumination with helicity eigenvalue $\lambda = \pm1$, the scattered field emitted by an antidual sample is of helicity $-\lambda$. Using Maxwell's equations and the helicity operator, such a situation can most generally be represented by the following constraints over the incident and scattered electromagnetic fields: $\sqrt{\varepsilon_m} \mathbf{E}_i = i\lambda \sqrt{\mu_m} \mathbf{H}_i$ and $\sqrt{\varepsilon_m}\mathbf{E}_s = -i\lambda \sqrt{\mu_m} \mathbf{H}_s$, where $\varepsilon_m$ and $\mu_m$ are the electric permittivity and magnetic permeability of the (lossless) medium in which the sample is embedded. Substituting these relations into the definition of the extinction component of the Poynting vector, it is obtained that $\mathbf{S}_{ext} = 0$ for any antidual scatterer. This implies that, regardless of the size, form, material constituents, or, even, the spatial shape of the incident illumination, an antidual scatterer produces a null extincted power. Thus, the relation among the absorbed, extincted and scattered powers enforces that, for an antidual sample, $W_a = -W_s$. As the scattered power is positive by definition, one is only left with two possibilities: either the scatterer is externally pumped with energy ($W_a < 0$) or both the absorbed and scattered powers must vanish. As a result, we conclude that passive antidual scatterers, i.e. those in which there is no external energy pumping, are precluded.

Let us underline the importance of this first contribution by briefly discussing previous works which are mainly devoted to the study of antidual dipolar spheres under plane-wave illumination \cite{Optimal3, Optimal31, Optimal32, CamaraExcep, AluEngheta, Optimal4, KerkerRef0, MiroZero, ZambranaKerker, DualAntidualModes, ShenGain, CrisAnapole, RAliGain}. In those works, it was essentially found that the zero forward scattering condition \cite{Optimal3, Optimal31, Optimal32, CamaraExcep, AluEngheta, Optimal4, KerkerRef0, MiroZero, ShenGain}, which follows from dealing with a cylindrically symmetric antidual particle \cite{ZambranaKerker, DualAntidualModes, ForBackCorbaton}, cannot be met according to the optical theorem. Indeed, the optical theorem states that the extincted power is proportional to the scattering amplitude in the forward direction \cite{AluEngheta, Bohren, Novotny,Jackson} and, thus, it is straightforward to deduce that $W_{ext} = 0$ for this type of scatterers. It was then concluded that Kerker's original second condition could not be achieved for passive dipolar spheres under plane-wave illumination. Now, our contribution generalizes this key result of linear electromagnetic scattering theory to an arbitrary antidual scatterer under general illumination conditions. Please note that in our derivation it is possible that the scattering amplitude has a non-vanishing value in the forward direction, which could be the case for non-cylindrical antidual particles. Our findings finally settle this long debate on the existence of passive antidual scatterers since forty years ago the second Kerker condition was formalized by Kerker, Wang and Giles until today.

Even though the construction of antidual scatterers with natural materials is not possible, we have found that there exists a condition which is not in contradiction with the energy conservation law and leads to scatterers that flip the helicity very efficiently.

\section{Helicity flipping passive scatterers}
We will be dealing with linear electromagnetic scatterers. Then,  as a starting point, we use the source-less time-independent Maxwell's equations:
\begin{align}
    \nonumber
    -i\omega\mathbf{D}(\mathbf{r}, \omega) &= \nabla\times\mathbf{H}(\mathbf{r}, \omega),~~~\nabla\cdot\mathbf{D}(\mathbf{r}, \omega) = 0,\\
    \label{Max}
    i\omega\mathbf{B}(\mathbf{r}, \omega) &= \nabla\times\mathbf{E}(\mathbf{r}, \omega),~~~\nabla\cdot\mathbf{B}(\mathbf{r}, \omega) = 0.
\end{align}
Let us consider, as a simplification of the problem, that the scattering process takes place upon an isotropic sample, i.e. $\mathbf{D}(\mathbf{r}, \omega) = \varepsilon(\mathbf{r})\mathbf{E}(\mathbf{r}, \omega)$ and $\mathbf{B}(\mathbf{r}, \omega) = \mu(\mathbf{r})\mathbf{H}(\mathbf{r}, \omega)$, where $\varepsilon(\mathbf{r})$ and $\mu(\mathbf{r})$ are the local electric permittivity and magnetic permeability, respectively. There is another formulation of Maxwell's equations, based on the the Riemann-Silberstein (RS) vector \cite{WeberRiemann, Silberstein, RoleRS}. Such a vector is defined as a linear combination of the electric and magnetic fields, i.e. $\mathbf{F}_\lambda (\mathbf{r}, \omega)= 2^{-1/2}[ \sqrt{\varepsilon(\mathbf{r})}\mathbf{E}(\mathbf{r}, \omega) +i\lambda \sqrt{\mu(\mathbf{r})} \mathbf{H}(\mathbf{r}, \omega)]$, with $\lambda = \pm 1$. This formulation is better for our purposes because it explicitly captures the polarization degree of freedom of electromagnetic waves. Actually, following the previous notation, the label $\lambda$ denotes the eigenvalue of the helicity operator $\Lambda$. When studying the properties of electromagnetic waves, it can be shown that the RS vector represents each of the two polarization degrees of freedom or helicities. This can be checked by employing the usual representation of the helicity operator, $\Lambda \rightarrow k^{-1}\nabla\times$, and expressing the time-independent Maxwell's equations in a homogeneous medium as a function the RS vector: $k^{-1}\nabla\times\mathbf{F}_\lambda = \lambda\mathbf{F}_\lambda$. Additionally, as we will see, this formulation of Maxwell's equations can be cast in terms of a hamiltonian notation, which will prove to be very useful in the next Section.

Let us rewrite Maxwell's equations given by \eqref{Max} in terms of the RS vector for inhomogeneous media \cite{Birula3, Birula1}:
\begin{align}
    \label{MaxRS1}
    \omega\mathbf{F}_+ = \frac{1}{\sqrt{n}}\nabla\times\left(\frac{\mathbf{F}_+}{\sqrt{n}}\right) + \frac{1}{n}\nabla\ln\sqrt{Z}\times\mathbf{F}_-\\
    \label{MaxRS2}
    \omega\mathbf{F}_- = -\frac{1}{\sqrt{n}}\nabla\times\left(\frac{\mathbf{F}_-}{\sqrt{n}}\right) - \frac{1}{n}\nabla\ln\sqrt{Z}\times\mathbf{F}_+\\[5pt]
    \label{MaxRS3}
    \nabla\cdot\mathbf{F}_+ = -\nabla\ln\sqrt{n}\cdot\mathbf{F_+} + \nabla\ln\sqrt{Z}\cdot\mathbf{F}_-\\[10pt]
    \label{MaxRS4}
    \nabla\cdot\mathbf{F}_- = -\nabla\ln\sqrt{n}\cdot\mathbf{F}_- + \nabla\ln\sqrt{Z}\cdot\mathbf{F}_+
\end{align}
where, for convenience, we have omitted the $(\mathbf{r}, \omega)$ dependencies and we have also defined the local impedance, $Z(\mathbf{r}) = \sqrt{\mu(\mathbf{r})/\varepsilon(\mathbf{r})}$, and refractive index, $n(\mathbf{r}) = \sqrt{\varepsilon(\mathbf{r})\mu(\mathbf{r})}$. Please note that Eqs. (\ref{MaxRS1})-(\ref{MaxRS4}) represent the time-independent Maxwell's equations in an isotropic and inhomogeneous medium exactly as those given by \eqref{Max}. In this form,  Faraday's and Ampère's laws given by \eqref{MaxRS1} and \eqref{MaxRS2} can be written as an eigenvalue problem:
\begin{align}
    \nonumber
    &H \begin{pmatrix} \mathbf{F}_+ \\ \mathbf{F}_- \end{pmatrix} = \omega \begin{pmatrix} \mathbf{F}_+\\ \mathbf{F}_- \end{pmatrix},~~\text{with}~~\\
    &H = \begin{pmatrix} \frac{1}{\sqrt{n}}\nabla\times\left(\frac{\cdot}{\sqrt{n}}\right) & \frac{1}{n}\nabla\ln \sqrt{Z}\times \\ -\frac{1}{n}\nabla\ln \sqrt{Z}\times & -\frac{1}{\sqrt{n}}\nabla\times\left(\frac{\cdot}{\sqrt{n}}\right) \end{pmatrix}.
    \label{H}
\end{align}
This particular way of expressing Faraday's and Ampère's laws has lead some authors to denote the operator $H$ as the "photon hamiltonian", in analogy with the time-independent Schrödinger's equation \cite{Birula3, Birula1, BarnettPhotonHam, IZBirula, PhotonHam1}. In this line, note that the frequency (or energy) of the system is completely determined by \eqref{H}. However, for our immediate purpose, \eqref{H} indicates that the role of a generic inhomogeneous medium is to interconnect the different helicity components of the electromagnetic field through the spatial derivatives of $Z$ and $n$.

The study of the impedance matching condition, i.e. media in which $\nabla Z = 0$, has been particularly fruitful for the communities of linear scattering theory and metamaterials \cite{Kerker, KerkerRef01, ACSSolomon, LatticeKerker, KerkerApp2, KivsharHuygens2, PfeifferGrbic2, KivsharHuygens}. Under this condition the duality symmetry is restored in macroscopic Maxwell's equations, leading to the conservation of helicity \cite{PRLMolina}. Indeed, in terms of Faraday-Ampère's laws given by \eqref{H}, it can be checked that the time-evolution of the helicity components are decoupled under this condition. Such a matching condition was first identified by Giles and Wild in the expressions of the Fresnel coefficients for a plane wave impinging on a plane surface \cite{GilesWild}. In particular, they found that when imposing the impedance matching condition, the reflection, $r$, and transmission, $t$, coefficients are the same for both $\text{s}$ and $\text{p}$ polarizations, i.e. $r_\text{s} = r_\text{p}$ and $t_\text{s} = t_\text{p}$, independently of the incidence angle. Following this pioneering work, Kerker, Wang and Giles one year later reported the effects of the impedance matching condition within Mie theory, i.e. in the problem of an electromagnetic wave scattering off a sphere \cite{Kerker}. Their findings confirmed that such a condition was very particular as it leads to specific analytical relations between the Mie coefficients, i.e. the amplitudes of the scattered electric/magnetic multipolar modes ($a_\ell$/$b_\ell$) and also the amplitudes of the internal electric/magnetic multipolar modes ($d_\ell$/$c_\ell$). In particular, it was found that the amplitude of the electric and magnetic multipolar modes, for both internal and scattered fields, is the same under this particular condition: $a_\ell = b_\ell \text{ and }d_\ell = c_\ell,~\forall~\ell$. Moreover, they realized that the analytical relation between the scattering coefficients implied an absence of backscattered fields. Since then, the impedance matching condition has most commonly been denoted as the first Kerker condition in the literature and adopts the analytical form $\varepsilon = \mu$ for a scatterer embedded in vacuum. 
\begin{table}[b]
\begin{center}
\begin{tabular}{ |c|c|c|c| } 
\hline
 & {\bf Fresnel} & {\bf Mie} \\
\hline
{\bf $\nabla Z = 0$} & $r_\text{s}/r_\text{p} = t_\text{s}/t_\text{p} = 1$ & $a_\ell/b_\ell = d_\ell/c_\ell = 1$ \\
 & &
\\ 
{\bf $\nabla n = 0$} & $~r_\text{s}/r_\text{p} = -t_\text{s}/t_\text{p} = -1~$ & $~a_\ell/b_\ell = -d_\ell/c_\ell \sim -1~$ \\
\hline
\end{tabular}
\caption{Fresnel and Mie coefficients under the impedance and refractive index matching conditions. $r_{\text{s}/\text{p}}$ and $t_{\text{s}/\text{p}}$ are Fresnel reflection and transmission coefficients for s and p polarized waves. $a_\ell$ and $b_\ell$ are the scattering Mie coefficients, whereas $d_\ell$ and $c_\ell$ are the internal Mie coefficients. Note that the last approximate equality only holds in the limit $s\rightarrow1$ and cannot be exactly fulfilled due to energy conservation.}
\label{table}
\end{center}
\end{table}
\noindent

While the role of the impedance has been analyzed in a wide variety of problems, the refractive index has been quite generally overlooked. Taking into account the time-evolution of the fields described by \eqref{H}, the index matching condition, i.e. media which fulfill $\nabla n = 0$, would imply having an environment in which the helicity components are mixed. At first sight, however, one is not able to tell the difference with a common dielectric medium ($\mu = 1$ and, thus, $Z = 1/n$) in which the helicity components are also coupled. The index matching condition was first reported by Giles and Wild in the expressions of the Fresnel coefficients \cite{GilesWild}. Under this condition they found that the reflection coefficients are changed by a relative sign regardless, again, of the incidence angle, i.e. $r_\text{s} = -r_\text{p}$ and $t_\text{s} = t_\text{p}$. Interestingly, this implies that, when scattering off an index-matched plane surface, the reflected light always has the opposite helicity \cite{Lakhtakia}. An attempt to study this condition within Mie theory was also made in Kerker's seminal paper but the authors were not so successful as they were with the analysis of the impedance matching effect. In fact, they exclusively focused on the study of the index matching condition for small spherical particles, overlooking the implications that this condition generally had on the analytical expressions of the Mie coefficients (see Section 4 in Ref. \cite{Kerker}). In what follows, we show that the refractive index matching condition does lead to spherical scatterers which very efficiently flip the helicity of electromagnetic waves, while still respecting the energy conservation law.
\begin{figure}[t]
    \centering
    \includegraphics[width = 0.4\textwidth]{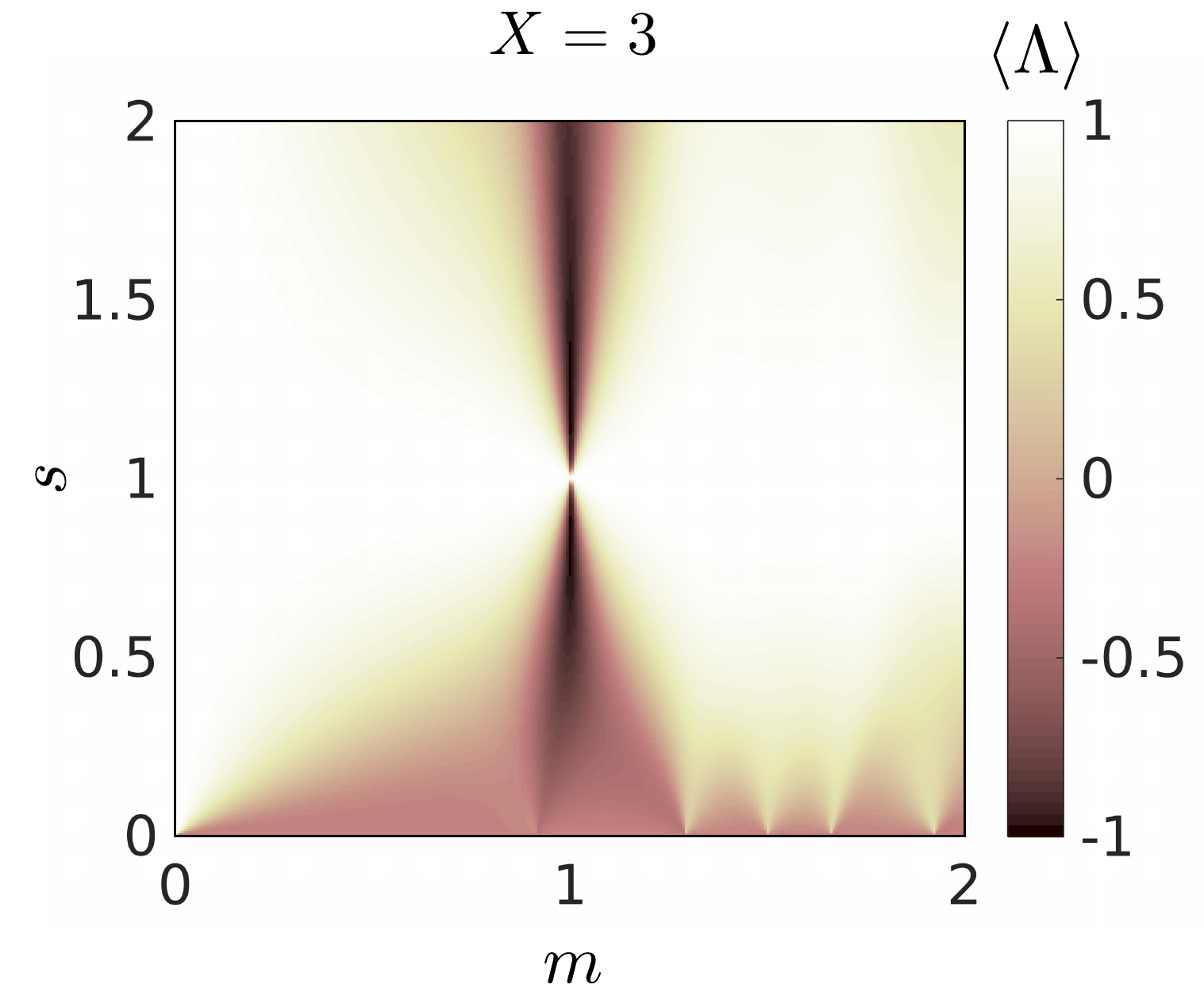}
    \caption{Helicity expectation value, $\langle \Lambda \rangle$, for a passive sphere of size parameter $X = 3$ illuminated by a circularly polarized plane wave. $\langle \Lambda \rangle$ is given as a function of the impedance contrast, $s = Z_1/Z_2$, and refractive index contrast, $m = n_1/n_2$. In particular,  $\langle \Lambda \rangle = 1$ implies a dual scatterer and $\langle \Lambda \rangle = -1$ an antidual one. The impedance matching condition is fulfilled through the line $s = 1$ and refractive index matching condition through the line $m = 1$.}
    \label{Hel}
\end{figure}

Whenever the refractive indices of the sphere ($n_1$) and the surrounding medium ($n_2$) are equal, we have found that the following analytical relation is fulfilled within the Mie coefficients: $a_\ell c_\ell = -b_\ell d_\ell,~\forall~\ell$. Furthermore, and most importantly, it can also be checked that the limit $\lim_{s \rightarrow 1} a_\ell = -b_\ell, ~\forall~\ell$ holds, where $s$ is the impedance contrast (see Supporting Information). This implies that, under the index matching condition ($n_1=n_2$), if one takes the impedances of the sphere ($Z_1$) and the surrounding medium ($Z_2$) to be increasingly similar ($s = Z_1/Z_2 \rightarrow 1$), then, the helicity of the scattered field is the opposite of the incident field. In other words, an antidual scatterer emerges \cite{ZambranaKerker}. However, note that such a limit can only be reached if there is no scatterer, as expected from our previous result on antiduality and energy conservation (see Section \ref{Sec2}). To shed light on this, we have computed the helicity expectation value, $\langle \Lambda \rangle$, for the fields scattered by a sphere of radius $a$ when illuminated by a circularly polarized plane wave. In Fig. \ref{Hel} we show the result in terms of the size parameter $X = k_0a$ of the sphere, where $k_0$ is the wavevector modulus of the incident wave in vacuum. The helicity expectation value is a bounded observable, $\langle \Lambda \rangle \in \left[-1, 1\right]$, and its extreme values imply a dual, $\langle \Lambda \rangle = 1$, or an antidual, $\langle \Lambda \rangle = -1$, scatterer \cite{Correlations, PRLJorge}. As it can be observed from Fig. \ref{Hel}, an outstanding behaviour is found when the index matching condition is fulfilled, through the line $m = 1$. Values of $\langle \Lambda \rangle \sim -1$ are obtained in the vicinities of such a line but never reaching the exact antiduality condition. Similar conclusions can be drawn from helicity maps for the large and the small sphere regime (see Supporting Information).

So, we have just shown that, even if passive antidual scatterers are precluded by the energy conservation law, one can obtain a pretty similar behaviour for index-matched spheres (see Fig. \ref{Hel}). This is based on the analytical expressions of the Mie coefficients under the refractive index matching condition, which had previously been overlooked in the literature. In addition, our analysis indicates that the impedance and refractive index matching conditions are paired in several contexts of electromagnetism. Indeed, we have explicitly shown that these two matching conditions have parallel implications over Fresnel and Mie coefficients (see Table \ref{table}). The pairing of the impedance and index matching conditions may be particularly useful in the field of Metamaterials, where one could in principle control the effective impedance and refractive index by properly engineering the inclusions and meta atoms. As it is known, helicity preserving and flipping surfaces emit, respectively, in the forward and backward directions provided that they have certain cylindrical symmetries \cite{ForBackCorbaton}. On the other hand, note that Fresnel coefficients do not depend on the angle of incidence for index-matched surfaces. Furthermore, reflected and transmitted powers are also independent of the helicity of the incident plane wave whenever $\nabla n = 0$. These properties could serve to construct angle- and polarization-independent ultrathin beam-splitters.

\section{Refractive index matching and the avoided level-crossing phenomenon}
The previous analysis points in the direction that the $\nabla n = 0$ condition allows for helicity flipping media as close as energy conservation allows. However, the exact mechanism for this helicity transfer still remains hidden. In this last part, we characterize the phenomenon that leads to the construction of efficient helicity flipping scatterers. To that aim, we evaluate the eigenfrequencies appearing in \eqref{H} for one-dimensional inhomogeneous media, i.e. systems in which $Z = Z(x)$ and $n = n(x)$. Due to the complexity of such a general problem, we have constrained to slowly varying environments for which time-independent perturbation theory may be applied. In other words, we analytically derive the form of Faraday's and Ampère's laws for smooth one-dimensional inhomogeneous media and compute the energies of the propagating electromagnetic waves in terms of $Z$ and $n$.

Standard time-independent perturbation theory may only be employed if the operator given by \eqref{H}, can be expressed as a sum of two terms, i.e. $H = H_0 + V$, where $H_0$ is the "hamiltonian" of a homogeneous medium and $V$ is the perturbation "potential" due to small inhomogeneities. Moreover, depending on whether we consider the perturbation in the direction of constant impedance or constant refractive index, $V$ has a different analytical form. Most generally, the perturbation potential in an arbitrary direction of the $(Z, n)$ parameter space can be constructed as a linear combination of the perturbation potentials in each direction. It can be shown that in the perturbative regime, the operator given by \eqref{H} breaks up into two terms: $H_0 = n^{-1}_0\bigl(\begin{smallmatrix}
		\nabla\times & 0\\
		0 & -\nabla\times
		\end{smallmatrix}\bigr)$
and $V = n^{-2}_0\bigl(\begin{smallmatrix}
		D & A\\
			-A & -D
		\end{smallmatrix}\bigr)$
where $n_0$ is a real constant representing the refractive index of the unperturbed system. The operators which define the components of the $V$ operator are $D = -[ik_te(x)\hat{u}_y + \delta_x\hat{u}_x]\times$ and $A = (2Z_0)^{-1}n_0 \partial_x{d}(x)\hat{u}_x\times$. We have defined the functions $d(x) = Z(x) - Z_0$ with $d(x)/Z_0 << 1$, where $Z_0$ is the impedance of the unperturbed system. Moreover, we have set $e(x) = n(x) - n_0$ with $e(x)/n_0 << 1$ and we have defined the operator $\delta_x = \sqrt{e(x)}\partial_x[\sqrt{e(x)}~\cdot~]$. Finally, without loss of generality, we consider that the solutions can be written as $\mathbf{\Psi}(\mathbf{r}) = \mathbf{\Psi}(x)e^{ik_t y}$ (see Supporting Information). Notice also that $V$ depends on the derivatives  of $e(x)$ and $d(x)$. Thus, for the perturbation theory to be valid, both $\partial_x e(x)/e(x)$ and $\partial_x d(x)/d(x)$ must be small compared to $k_t$. This precludes the use of sharp and step-like functions.

Once we have obtained the complete perturbation potential, we need a set of exact solutions of the unperturbed system. Such a set of solutions is obtained by solving the system of coupled differential equations $H_0\mathbf{\Psi}(x) = \omega\mathbf{\Psi}(x)$. As expected, the solutions are circularly polarized plane waves propagating in the $XY$ plane. We find four different types of normalized eigenstates: $\mathbf{\Psi}_{\lambda\pm}(x)=\sqrt{\omega} \hat{u}_\lambda \exp(\pm i k_x x)$, which are classified by their helicity, $\lambda$, and direction of propagation, $\pm k_x$, with $k_x^2 = \omega^2n_0^2 - k_t^2$; and $\hat{u}_\lambda$ is a normalized polarization vector. All of these eigenvectors fulfill $H_0\mathbf{\Psi}_{\lambda\pm}(x) = \omega\mathbf{\Psi}_{\lambda\pm}(x)$, implying that they are degenerate (see Supporting Information). This means that the exact solutions of Maxwell's equations in a homogeneous medium share the same energy and, thus, degenerate time-independent perturbation theory must be employed. In our case, the approximate solutions to the perturbed system, $\mathbf{\Psi}^{(1)}$, can be found as linear superpositions of the unperturbed system eigenstates ($\mathbf{\Psi}^{(1)}=\sum_{j} K_j \mathbf{\Psi}_j$) by diagonalizing the $4\times 4$ matrix constructed from the matrix elements $\braket{\mathbf{\Psi}_j|V\mathbf{\Psi}_i}$ where in $\mathbf{\Psi}_k$, $k\in\{++,+-,-+,--\}$. Finally, the eigenvalues of the matrix correspond to the first order energy corrections.

As the problem we are interested in is one dimensional, the matrix elements are computed as $\braket{\mathbf{\Psi}_j|V\mathbf{\Psi}_i} = \int_{-\infty}^{\infty}dx~[\mathbf{\Psi}_j(x)]^*\cdot [V\mathbf{\Psi}_i(x)]$. The computation of such integrals results in the matrix equation M$\mathbf{K} = \frac{\Delta \omega}{\omega} \mathbf{K}$, where $\mathbf{K} = (K_1, K_2, K_3, K_4)^T$. In addition, $\Delta \omega$ represents the first order correction to the energy such that the perturbed energies are obtained as $\omega^{(1)} = \omega + \Delta \omega$. After a similarity transformation, the matrix M can be written in a form M$'$ that better allows us to identify the physical properties of the problem. Thus, we are left with the following eigensystem:  M$'\mathbf{K}' = \frac{\Delta \omega}{\omega} \mathbf{K}'$, where
\begin{equation}
    \label{PertW3}
    \text{M}'
    =
    \begin{pmatrix}
    \mathbb{D} + |\mathbb{F}| & \mathbb{C}^* & 0 & 0 \\
    \mathbb{C} & -(\mathbb{D} - |\mathbb{F}|) & 0 & 0 \\
    0 & 0 & \mathbb{D} - |\mathbb{F}| & \mathbb{C}^* \\
    0 & 0 & \mathbb{C} & -(\mathbb{D} + |\mathbb{F}|)
    \end{pmatrix}
\end{equation}
and $\mathbf{K}' = (K'_1, K'_4, K'_2, K'_3)^T$. Please note that the helicity components are reordered within the vector $\mathbf{K}'$. In the new basis, $K'_1$ and $K'_2$ still represent positive helicity components, and also, the negative helicity components are still associated with $K'_3$ and $K'_4$ (see Supporting Information). Parameter $\mathbb{C}$ in \eqref{PertW3} captures the effect of the inhomogeneities in the impedance, i.e. $\mathbb{C}$ vanishes whenever $\partial_x Z(x) = 0$. On the other hand, $\mathbb{D}$ and $\mathbb{F}$ capture the effect of the inhomogeneities in the refractive index, i.e. $\mathbb{D}$ and $\mathbb{F}$ are zero whenever $\partial_x n(x) = 0$. 

Bearing this in mind, we can straightforwardly analyze the effect of inhomogeneities from the elements of the matrix in \eqref{PertW3}. First of all, it is clear that for an arbitrary perturbation in the $(Z, n)$ parameter space, the whole system can be split into two independent subsystems. Moreover, due to their dimensionality, each of the subsystems can be analyzed in close analogy to a quantum mechanical two-level system. In particular, the case of impedance matching ($\mathbb{C} = 0$) implies, in this formalism, that the levels are decoupled and, thus, there is no mixing between the helicities, whereas the refractive index matching condition ($\mathbb{F} = \mathbb{D} = 0$) makes the two-level systems be simultaneously degenerate and coupled. On the other hand, the expression of the energy corrections is
$\Delta \omega = \pm \omega(\mathbb{S} \pm |\mathbb{F}|)$,
where $\mathbb{S} = \sqrt{\mathbb{D}^2 + |\mathbb{C}|^2}$, and for index-matched inhomogeneous media the perturbed energies are computed as $\omega^{(1)} = \omega \pm \omega |\mathbb{C}|$. This is the exact mathematical form that energies take in a resonant two-level system. In addition, whenever $n(x)$ is a constant function, perturbed states can be written as linear superpositions of the unperturbed states with amplitudes of the same modulus, which is also a key characteristic of the systems undergoing quantum resonance \cite{CohenTannoudji}.
\begin{figure}[t]
    \centering
    \includegraphics[width = 0.45\textwidth]{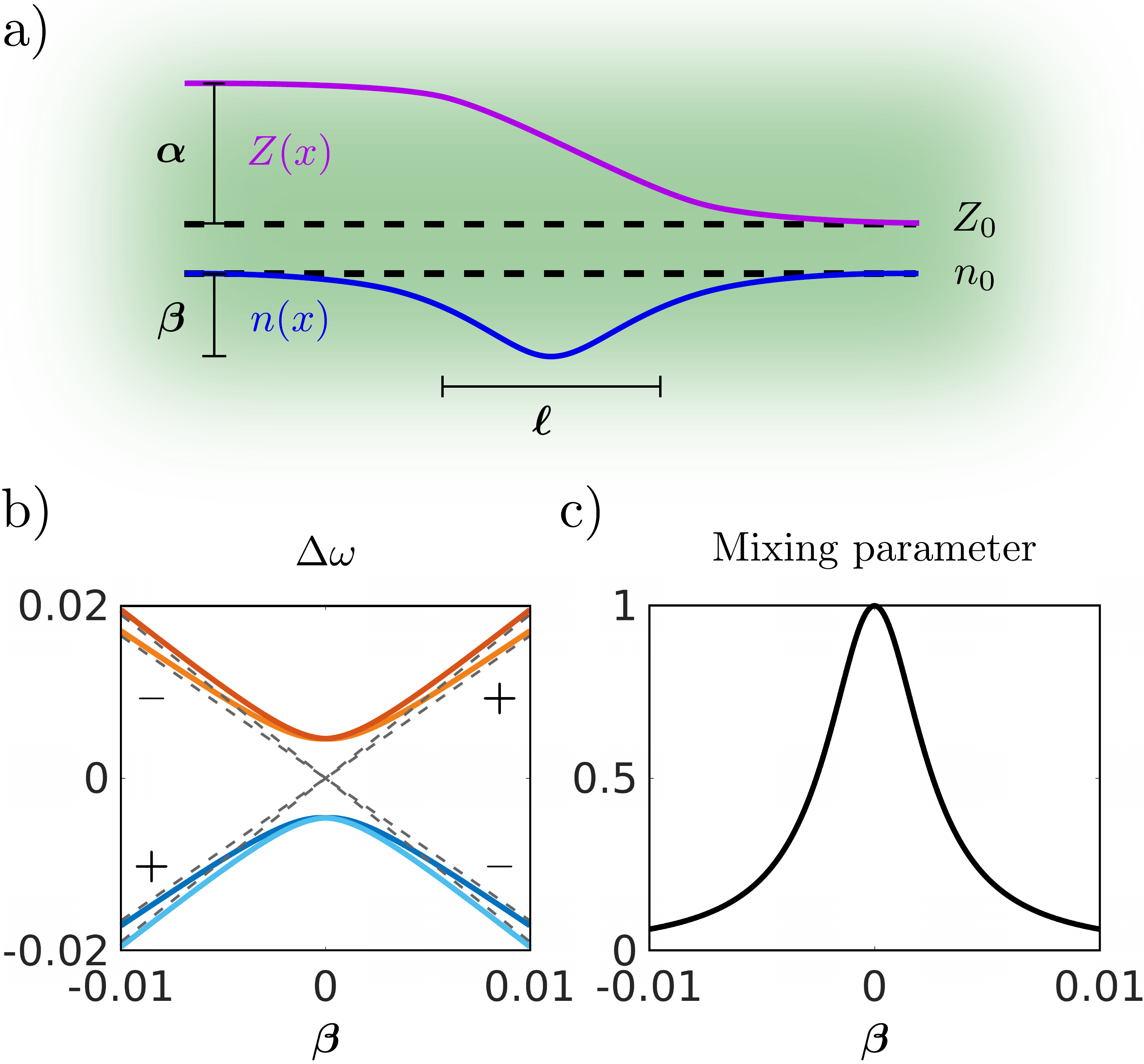}
    \caption{a) Sketch of an inhomogeneous medium with $Z(x) = Z_0 + d(x)$ and $n(x) = n_0 + e(x)$, with the functions $d(x) = \alpha/[1 + \exp{(x/\ell)}]$ and $e(x) = -\beta\exp{(-x^2/\ell^2)}$. b) First order energy corrections: $Z_0 = n_0 = 1$, $\omega = 1$, $k_t  = 0.4$, $\ell = 1$ and $\alpha = 0.01$ (in colors). All units are normalized to an arbitrary frequency $\omega_0$ and $c=1$. The grey dashed lines are obtained with the same parameters but fixing $\alpha = 0$, which implies that $\mathbb{C} = 0$ in this case. c) Mixing parameter of the two-level systems \cite{CohenTannoudji}. Whenever $\beta = 0$, the refractive index matching condition is fulfilled.}
    \label{Anticrossing}
\end{figure}

In Fig. \ref{Anticrossing} we have numerically computed a particular case of a one-dimensional inhomogeneous system. We have considered that the perturbation is given by the functions $d(x) = \alpha/[1 + \exp{(x/\ell)}]$ and $e(x) = -\beta\exp{(-x^2/\ell^2)}$ (see Fig. \ref{Anticrossing}a). The amplitude of the perturbation in the impedance is determined by the parameter $\alpha$ and in the refractive index by $\beta$. Note that whenever $\alpha = 0$ helicity is preserved and, on the other hand, whenever $\beta = 0$ the refractive index matching condition is fulfilled. To meet the requirements imposed by perturbation theory we have chosen $\alpha = 0.01 \ll Z_0$ and $|\beta| \le 0.01 \ll n_0$. In Fig. \ref{Anticrossing}b we have computed the corrections to the energy as a function of $\beta$ (in colors). For comparison, in the same figure, we also display the energy corrections for the case in which $\alpha = 0$ (grey dashed lines). Whenever $\alpha = 0$ we have that $\mathbb{C} = 0$ and, thus, the helicity components are decoupled in this case. From the form of the matrix in \eqref{PertW3} it can be checked that the ascending (descending) diagonal grey lines correspond to solutions with well-defined positive (negative) helicity. Finally, in Fig. \ref{Anticrossing}c, we have computed the mixing parameters of the two-level systems identified in \eqref{PertW3}, which happen to be the same. Among others, this parameter has been employed to study the resonant mixing of neutrino flavours in matter and it is a signature of quantum resonance when it approaches unity \cite{MSW1, MSW2}. The reported results are completely compatible with the phenomenon of the avoided level-crossing.

To sum up, our results indicate that index-matched media induce a resonant helicity mixing on the electromagnetic waves that propagate through them. We have reached to this conclusion by applying perturbation theory to Maxwell's equations for smooth varying environments and computing the energies of the propagating waves in terms of the medium parameters. Strikingly and completely unexpectedly, we have identified that the refractive index matching condition ($\nabla n = 0$) leads to the phenomenon of avoided level-crossing (see Fig. \ref{Anticrossing}). A great deal of phenomena in Physics are related to the avoided crossing of energy levels, but this is the first time that such a connection is made for the helicity components of electromagnetic fields. In our view, there is an effect which closely resembles the resonant helicity mixing condition: the Mikheyev–Smirnov–Wolfenstein (MSW) effect in neutrino physics \cite{MSW0, MSW1}. This effect, which has been regarded as the solution to the solar neutrino problem \cite{MSWSolar}, takes place when neutrinos propagate through a medium with varying matter density, $\rho(x)$. Similarly, there also exists a condition over the medium for which neutrino flavours are resonantly mixed, i.e. whenever $\rho(x) = \rho_R$. However, the resonant density, $\rho_R$, is a function of the initial neutrino energy and, thus, only neutrinos that are released with enough energy hit the resonance in their way out of the sun. This turns out on an almost complete conversion of the high energy electron neutrinos generated by the sun into other neutrino flavours. One can observe the mathematical concordance of the analogy between neutrino flavours and electromagnetic helicity components, and the density matching and refractive index matching conditions. 

\section{Conclusions}
In conclusion, we have shown that passive antidual scatterers are precluded by the energy conservation law in linear electromagnetic scattering theory. The result is completely general and it is independent of the size, shape, material constituent or, even, the spatial form of the illuminating field. Then, we have presented the condition of refractive index matching and showed that it appears paired with the impedance matching condition in several contexts of electromagnetism including Fresnel and Mie coefficients. Evaluating the analytical form of Mie coefficients under the index matching condition, something which had been previously overlooked, we have found that this condition leads to spherical scatterers which flip the helicity of the incoming light very efficiently. Importantly, these scatterers do not contradict the energy conservation law and, thus, could in principle be constructed from natural materials or metamaterials. Finally, we have concluded that, index-matched media induce a resonant helicity mixing on the electromagnetic waves that propagate through them. We have reached to this conclusion by evaluating the energies of propagating waves in terms of the medium parameters and identifying that the index matching condition leads to the phenomenon of avoided level-crossing.

We believe that our findings will have an impact on, at least, three different communities. Our first contribution generalizes a key result of linear electromagnetic scattering theory to generic passive scatterers under general illumination conditions. This closes a historical discussion on the zero forward scattering problem \cite{Optimal3, Optimal31, Optimal32, Optimal4, AluEngheta, KerkerRef0, CamaraExcep, MiroZero, ShenGain}, the second Kerker condition \cite{AluEngheta, NietoVesperinasK2, MiroshisenkoK2, MiroZero, PRROlmos, PRLJorge, Optimal4, ShenGain} and the possible existence of passive antidual scatterers \cite{ZambranaKerker, DualAntidualModes, ForBackCorbaton, PRROlmos, PRLJorge, RAliGain}. Moreover, identifying the refractive index matching condition as being paired to the impedance matching condition in several contexts of electromagnetism opens up a great deal of new possibilities in the field of Metamaterials. Exactly as impedance-matched meta atoms are commonly employed to maximize transmissivity \cite{KerkerApp2, KivsharHuygens2, PfeifferGrbic2, KivsharHuygens}, index-matched particles are good candidates to build reflective photonic devices. Also, index-matched metasurfaces could serve to construct angle- and polarization-independent ultrathin beam-splitters. Finally, the connection with quantum two level-systems aligns the resonant helicity mixing condition with other phenomena such as the MSW effect. This identification opens up a window of possible cross-pollination of ideas with Particle Physics.

\begin{acknowledgements}
J.L.A. acknowledges the warm hospitality that Iwo Bialynicki-Birula and Zofia Bialynicka-Birula have extended to him in Warsaw where part of this work has been developed. J.L.A., J.O.T., C. D. and A.G.E acknowledge support from Project No. PID2019-109905GA-C22 of the Spanish Ministerio de Ciencia, Innovaci\'on y Universidades (MICIU), IKUR Strategy under the collaboration agreement between Ikerbasque Foundation and DIPC on behalf of the Department of Education of the Basque Government, the Basque Government Elkartek program (KK-2021/00082), Programa de ayudas de apoyo a los agentes de la Red Vasca de Ciencia, Tecnolog\'ia e Innovaci\'on acreditados en la categor\'ia de Centros de Investigaci\'on B\'asica y de Excelencia (Programa BERC) from the Departamento de Universidades e Investigaci\'on del Gobierno Vasco and Centros Severo Ochoa AEI/CEX2018-000867-S from the Spanish Ministerio de Ciencia e Innovaci\'on. J.O.T. acknowledges support from the Juan de la Cierva fellowship No. FJC2021-047090-I of  MCIN/AEI/10.13039/501100011033 and NextGenerationEU/PRTR. G.M.T received funding from the IKUR Strategy under the collaboration agreement between Ikerbasque Foundation and DIPC/MPC on behalf of the Department of Education of the Basque Government. P. H. acknowledges support from the DIPC visitor program.
\end{acknowledgements}

\clearpage
\bibliography{mybib}

\begin{thebibliography}{63}%
\makeatletter
\providecommand \@ifxundefined [1]{%
 \@ifx{#1\undefined}
}%
\providecommand \@ifnum [1]{%
 \ifnum #1\expandafter \@firstoftwo
 \else \expandafter \@secondoftwo
 \fi
}%
\providecommand \@ifx [1]{%
 \ifx #1\expandafter \@firstoftwo
 \else \expandafter \@secondoftwo
 \fi
}%
\providecommand \natexlab [1]{#1}%
\providecommand \enquote  [1]{``#1''}%
\providecommand \bibnamefont  [1]{#1}%
\providecommand \bibfnamefont [1]{#1}%
\providecommand \citenamefont [1]{#1}%
\providecommand \href@noop [0]{\@secondoftwo}%
\providecommand \href [0]{\begingroup \@sanitize@url \@href}%
\providecommand \@href[1]{\@@startlink{#1}\@@href}%
\providecommand \@@href[1]{\endgroup#1\@@endlink}%
\providecommand \@sanitize@url [0]{\catcode `\\12\catcode `\$12\catcode
  `\&12\catcode `\#12\catcode `\^12\catcode `\_12\catcode `\%12\relax}%
\providecommand \@@startlink[1]{}%
\providecommand \@@endlink[0]{}%
\providecommand \url  [0]{\begingroup\@sanitize@url \@url }%
\providecommand \@url [1]{\endgroup\@href {#1}{\urlprefix }}%
\providecommand \urlprefix  [0]{URL }%
\providecommand \Eprint [0]{\href }%
\providecommand \doibase [0]{http://dx.doi.org/}%
\providecommand \selectlanguage [0]{\@gobble}%
\providecommand \bibinfo  [0]{\@secondoftwo}%
\providecommand \bibfield  [0]{\@secondoftwo}%
\providecommand \translation [1]{[#1]}%
\providecommand \BibitemOpen [0]{}%
\providecommand \bibitemStop [0]{}%
\providecommand \bibitemNoStop [0]{.\EOS\space}%
\providecommand \EOS [0]{\spacefactor3000\relax}%
\providecommand \BibitemShut  [1]{\csname bibitem#1\endcsname}%
\let\auto@bib@innerbib\@empty
%</preamble>
\bibitem [{\citenamefont {Allen}\ and\ \citenamefont
  {Eberly}(1975)}]{Allen_Eberly}%
  \BibitemOpen
  \bibfield  {author} {\bibinfo {author} {\bibfnamefont {L.}~\bibnamefont
  {Allen}}\ and\ \bibinfo {author} {\bibfnamefont {J.~H.}\ \bibnamefont
  {Eberly}},\ }\href@noop {} {\emph {\bibinfo {title} {Optical resonance and
  two-level atoms}}}\ (\bibinfo  {publisher} {John Wiley and Sons, Inc.},\
  \bibinfo {year} {1975})\BibitemShut {NoStop}%
\bibitem [{\citenamefont {Novotny}(2010)}]{Novotny_StrongCoupling}%
  \BibitemOpen
  \bibfield  {author} {\bibinfo {author} {\bibfnamefont {L.}~\bibnamefont
  {Novotny}},\ }\href@noop {} {\bibfield  {journal} {\bibinfo  {journal}
  {American Journal of Physics}\ }\textbf {\bibinfo {volume} {78}},\ \bibinfo
  {pages} {1199} (\bibinfo {year} {2010})}\BibitemShut {NoStop}%
\bibitem [{\citenamefont {Tame}\ \emph {et~al.}(2013)\citenamefont {Tame},
  \citenamefont {McEnery}, \citenamefont {Ozdemir}, \citenamefont {Lee},
  \citenamefont {Maier},\ and\ \citenamefont {Kim}}]{StrongCoupling1}%
  \BibitemOpen
  \bibfield  {author} {\bibinfo {author} {\bibfnamefont {M.~S.}\ \bibnamefont
  {Tame}}, \bibinfo {author} {\bibfnamefont {K.~R.}\ \bibnamefont {McEnery}},
  \bibinfo {author} {\bibfnamefont {S.~K.}\ \bibnamefont {Ozdemir}}, \bibinfo
  {author} {\bibfnamefont {J.}~\bibnamefont {Lee}}, \bibinfo {author}
  {\bibfnamefont {S.~A.}\ \bibnamefont {Maier}}, \ and\ \bibinfo {author}
  {\bibfnamefont {M.~S.}\ \bibnamefont {Kim}},\ }\href@noop {} {\bibfield
  {journal} {\bibinfo  {journal} {Nature Physics}\ }\textbf {\bibinfo {volume}
  {9}},\ \bibinfo {pages} {329} (\bibinfo {year} {2013})}\BibitemShut {NoStop}%
\bibitem [{\citenamefont {Chikkaraddy}\ \emph {et~al.}(2016)\citenamefont
  {Chikkaraddy}, \citenamefont {de~Nijs}, \citenamefont {Benz}, \citenamefont
  {Barrow}, \citenamefont {Scherman}, \citenamefont {Rosta}, \citenamefont
  {Demetriadou}, \citenamefont {Fox}, \citenamefont {Hess},\ and\ \citenamefont
  {Baumberg}}]{StrongCoupling2}%
  \BibitemOpen
  \bibfield  {author} {\bibinfo {author} {\bibfnamefont {R.}~\bibnamefont
  {Chikkaraddy}}, \bibinfo {author} {\bibfnamefont {B.}~\bibnamefont
  {de~Nijs}}, \bibinfo {author} {\bibfnamefont {F.}~\bibnamefont {Benz}},
  \bibinfo {author} {\bibfnamefont {S.~J.}\ \bibnamefont {Barrow}}, \bibinfo
  {author} {\bibfnamefont {O.~A.}\ \bibnamefont {Scherman}}, \bibinfo {author}
  {\bibfnamefont {E.}~\bibnamefont {Rosta}}, \bibinfo {author} {\bibfnamefont
  {A.}~\bibnamefont {Demetriadou}}, \bibinfo {author} {\bibfnamefont
  {P.}~\bibnamefont {Fox}}, \bibinfo {author} {\bibfnamefont {O.}~\bibnamefont
  {Hess}}, \ and\ \bibinfo {author} {\bibfnamefont {J.~J.}\ \bibnamefont
  {Baumberg}},\ }\href@noop {} {\bibfield  {journal} {\bibinfo  {journal}
  {Nature}\ }\textbf {\bibinfo {volume} {535}},\ \bibinfo {pages} {127}
  (\bibinfo {year} {2016})}\BibitemShut {NoStop}%
\bibitem [{\citenamefont {Lukin}\ \emph {et~al.}(1999)\citenamefont {Lukin},
  \citenamefont {Yelin}, \citenamefont {Fleischhauer},\ and\ \citenamefont
  {Scully}}]{Tripod1}%
  \BibitemOpen
  \bibfield  {author} {\bibinfo {author} {\bibfnamefont {M.~D.}\ \bibnamefont
  {Lukin}}, \bibinfo {author} {\bibfnamefont {S.~F.}\ \bibnamefont {Yelin}},
  \bibinfo {author} {\bibfnamefont {M.}~\bibnamefont {Fleischhauer}}, \ and\
  \bibinfo {author} {\bibfnamefont {M.~O.}\ \bibnamefont {Scully}},\
  }\href@noop {} {\bibfield  {journal} {\bibinfo  {journal} {Phys. Rev. A}\
  }\textbf {\bibinfo {volume} {60}},\ \bibinfo {pages} {3225} (\bibinfo {year}
  {1999})}\BibitemShut {NoStop}%
\bibitem [{\citenamefont {Vewinger}\ \emph {et~al.}(2003)\citenamefont
  {Vewinger}, \citenamefont {Heinz}, \citenamefont {Garcia~Fernandez},
  \citenamefont {Vitanov},\ and\ \citenamefont {Bergmann}}]{Tripod2}%
  \BibitemOpen
  \bibfield  {author} {\bibinfo {author} {\bibfnamefont {F.}~\bibnamefont
  {Vewinger}}, \bibinfo {author} {\bibfnamefont {M.}~\bibnamefont {Heinz}},
  \bibinfo {author} {\bibfnamefont {R.}~\bibnamefont {Garcia~Fernandez}},
  \bibinfo {author} {\bibfnamefont {N.~V.}\ \bibnamefont {Vitanov}}, \ and\
  \bibinfo {author} {\bibfnamefont {K.}~\bibnamefont {Bergmann}},\ }\href@noop
  {} {\bibfield  {journal} {\bibinfo  {journal} {Phys. Rev. Lett.}\ }\textbf
  {\bibinfo {volume} {91}},\ \bibinfo {pages} {213001} (\bibinfo {year}
  {2003})}\BibitemShut {NoStop}%
\bibitem [{\citenamefont {Wolfenstein}(1978)}]{MSW0}%
  \BibitemOpen
  \bibfield  {author} {\bibinfo {author} {\bibfnamefont {L.}~\bibnamefont
  {Wolfenstein}},\ }\href@noop {} {\bibfield  {journal} {\bibinfo  {journal}
  {Phys. Rev. D}\ }\textbf {\bibinfo {volume} {17}},\ \bibinfo {pages} {2369}
  (\bibinfo {year} {1978})}\BibitemShut {NoStop}%
\bibitem [{\citenamefont {Mikheyev}\ and\ \citenamefont
  {Smirnov}(1986)}]{MSW1}%
  \BibitemOpen
  \bibfield  {author} {\bibinfo {author} {\bibfnamefont {S.~P.}\ \bibnamefont
  {Mikheyev}}\ and\ \bibinfo {author} {\bibfnamefont {A.~Y.}\ \bibnamefont
  {Smirnov}},\ }\href@noop {} {\bibfield  {journal} {\bibinfo  {journal} {Il
  Nuovo Cimento C}\ }\textbf {\bibinfo {volume} {9}},\ \bibinfo {pages} {17}
  (\bibinfo {year} {1986})}\BibitemShut {NoStop}%
\bibitem [{\citenamefont {Mikheyev}\ and\ \citenamefont
  {Smirnov}(1989)}]{MSW2}%
  \BibitemOpen
  \bibfield  {author} {\bibinfo {author} {\bibfnamefont {S.~P.}\ \bibnamefont
  {Mikheyev}}\ and\ \bibinfo {author} {\bibfnamefont {A.~Y.}\ \bibnamefont
  {Smirnov}},\ }\href@noop {} {\bibfield  {journal} {\bibinfo  {journal}
  {Progress in Particle and Nuclear Physics}\ }\textbf {\bibinfo {volume}
  {23}},\ \bibinfo {pages} {41} (\bibinfo {year} {1989})}\BibitemShut {NoStop}%
\bibitem [{\citenamefont {Bialynicki-Birula}(1994)}]{Birula3}%
  \BibitemOpen
  \bibfield  {author} {\bibinfo {author} {\bibfnamefont {I.}~\bibnamefont
  {Bialynicki-Birula}},\ }\href@noop {} {\bibfield  {journal} {\bibinfo
  {journal} {Acta Physica Polonica A}\ }\textbf {\bibinfo {volume} {86}},\
  \bibinfo {pages} {97} (\bibinfo {year} {1994})}\BibitemShut {NoStop}%
\bibitem [{\citenamefont {Bialynicki-Birula}(1996)}]{Birula1}%
  \BibitemOpen
  \bibfield  {author} {\bibinfo {author} {\bibfnamefont {I.}~\bibnamefont
  {Bialynicki-Birula}},\ }in\ \href@noop {} {\emph {\bibinfo {booktitle}
  {Progress in Optics}}},\ Vol.~\bibinfo {volume} {36},\ \bibinfo {editor}
  {edited by\ \bibinfo {editor} {\bibfnamefont {E.}~\bibnamefont {Wolf}}}\
  (\bibinfo  {publisher} {Elsevier},\ \bibinfo {year} {1996})\ Chap.~\bibinfo
  {chapter} {5}, pp.\ \bibinfo {pages} {245--294}\BibitemShut {NoStop}%
\bibitem [{\citenamefont {Messiah}(1962)}]{Messiah}%
  \BibitemOpen
  \bibfield  {author} {\bibinfo {author} {\bibfnamefont {A.}~\bibnamefont
  {Messiah}},\ }\href@noop {} {\emph {\bibinfo {title} {Quantum Mechanics,
  Volume 2.}}}\ (\bibinfo  {publisher} {North Holland Publishing Company},\
  \bibinfo {year} {1962})\BibitemShut {NoStop}%
\bibitem [{\citenamefont {Fern\'andez-Corbaton}\ \emph
  {et~al.}(2013)\citenamefont {Fern\'andez-Corbaton}, \citenamefont
  {Zambrana-Puyalto}, \citenamefont {Tischler}, \citenamefont {Vidal},
  \citenamefont {Juan},\ and\ \citenamefont {Molina-Terriza}}]{PRLMolina}%
  \BibitemOpen
  \bibfield  {author} {\bibinfo {author} {\bibfnamefont {I.}~\bibnamefont
  {Fern\'andez-Corbaton}}, \bibinfo {author} {\bibfnamefont {X.}~\bibnamefont
  {Zambrana-Puyalto}}, \bibinfo {author} {\bibfnamefont {N.}~\bibnamefont
  {Tischler}}, \bibinfo {author} {\bibfnamefont {X.}~\bibnamefont {Vidal}},
  \bibinfo {author} {\bibfnamefont {M.~L.}\ \bibnamefont {Juan}}, \ and\
  \bibinfo {author} {\bibfnamefont {G.}~\bibnamefont {Molina-Terriza}},\
  }\href@noop {} {\bibfield  {journal} {\bibinfo  {journal} {Physical Review
  Letters}\ }\textbf {\bibinfo {volume} {111}},\ \bibinfo {pages} {060401}
  (\bibinfo {year} {2013})}\BibitemShut {NoStop}%
\bibitem [{\citenamefont {Bliokh}\ \emph {et~al.}(2019)\citenamefont {Bliokh},
  \citenamefont {Leykam}, \citenamefont {Lein},\ and\ \citenamefont
  {Nori}}]{BliokhHel}%
  \BibitemOpen
  \bibfield  {author} {\bibinfo {author} {\bibfnamefont {K.~Y.}\ \bibnamefont
  {Bliokh}}, \bibinfo {author} {\bibfnamefont {D.}~\bibnamefont {Leykam}},
  \bibinfo {author} {\bibfnamefont {M.}~\bibnamefont {Lein}}, \ and\ \bibinfo
  {author} {\bibfnamefont {F.}~\bibnamefont {Nori}},\ }\href@noop {} {\bibfield
   {journal} {\bibinfo  {journal} {Nature Communications}\ }\textbf {\bibinfo
  {volume} {10}},\ \bibinfo {pages} {580} (\bibinfo {year} {2019})}\BibitemShut
  {NoStop}%
\bibitem [{\citenamefont {Person}\ \emph {et~al.}(2013)\citenamefont {Person},
  \citenamefont {Jain}, \citenamefont {Lapin}, \citenamefont {S{\'a}enz},
  \citenamefont {Wicks},\ and\ \citenamefont {Novotny}}]{KerkerRef01}%
  \BibitemOpen
  \bibfield  {author} {\bibinfo {author} {\bibfnamefont {S.}~\bibnamefont
  {Person}}, \bibinfo {author} {\bibfnamefont {M.}~\bibnamefont {Jain}},
  \bibinfo {author} {\bibfnamefont {Z.}~\bibnamefont {Lapin}}, \bibinfo
  {author} {\bibfnamefont {J.~J.}\ \bibnamefont {S{\'a}enz}}, \bibinfo {author}
  {\bibfnamefont {G.}~\bibnamefont {Wicks}}, \ and\ \bibinfo {author}
  {\bibfnamefont {L.}~\bibnamefont {Novotny}},\ }\href@noop {} {\bibfield
  {journal} {\bibinfo  {journal} {Nano Letters}\ }\textbf {\bibinfo {volume}
  {13}},\ \bibinfo {pages} {1806} (\bibinfo {year} {2013})}\BibitemShut
  {NoStop}%
\bibitem [{\citenamefont {Fu}\ \emph {et~al.}(2013)\citenamefont {Fu},
  \citenamefont {Kuznetsov}, \citenamefont {Miroshnichenko}, \citenamefont
  {Yu},\ and\ \citenamefont {Luk’yanchuk}}]{KerkerRef1}%
  \BibitemOpen
  \bibfield  {author} {\bibinfo {author} {\bibfnamefont {Y.~H.}\ \bibnamefont
  {Fu}}, \bibinfo {author} {\bibfnamefont {A.~I.}\ \bibnamefont {Kuznetsov}},
  \bibinfo {author} {\bibfnamefont {A.~E.}\ \bibnamefont {Miroshnichenko}},
  \bibinfo {author} {\bibfnamefont {Y.~F.}\ \bibnamefont {Yu}}, \ and\ \bibinfo
  {author} {\bibfnamefont {B.}~\bibnamefont {Luk’yanchuk}},\ }\href@noop {}
  {\bibfield  {journal} {\bibinfo  {journal} {Nature Communications}\ }\textbf
  {\bibinfo {volume} {4}} (\bibinfo {year} {2013})}\BibitemShut {NoStop}%
\bibitem [{\citenamefont {Staude}\ \emph
  {et~al.}(2013{\natexlab{a}})\citenamefont {Staude}, \citenamefont
  {Miroshnichenko}, \citenamefont {Decker}, \citenamefont {Fofang},
  \citenamefont {Liu}, \citenamefont {Gonzales}, \citenamefont {Dominguez},
  \citenamefont {Luk}, \citenamefont {Neshev}, \citenamefont {Brener},\ and\
  \citenamefont {Kivshar}}]{KerkerRef2}%
  \BibitemOpen
  \bibfield  {author} {\bibinfo {author} {\bibfnamefont {I.}~\bibnamefont
  {Staude}}, \bibinfo {author} {\bibfnamefont {A.~E.}\ \bibnamefont
  {Miroshnichenko}}, \bibinfo {author} {\bibfnamefont {M.}~\bibnamefont
  {Decker}}, \bibinfo {author} {\bibfnamefont {N.~T.}\ \bibnamefont {Fofang}},
  \bibinfo {author} {\bibfnamefont {S.}~\bibnamefont {Liu}}, \bibinfo {author}
  {\bibfnamefont {E.}~\bibnamefont {Gonzales}}, \bibinfo {author}
  {\bibfnamefont {J.}~\bibnamefont {Dominguez}}, \bibinfo {author}
  {\bibfnamefont {T.~S.}\ \bibnamefont {Luk}}, \bibinfo {author} {\bibfnamefont
  {D.~N.}\ \bibnamefont {Neshev}}, \bibinfo {author} {\bibfnamefont
  {I.}~\bibnamefont {Brener}}, \ and\ \bibinfo {author} {\bibfnamefont
  {Y.}~\bibnamefont {Kivshar}},\ }\href@noop {} {\bibfield  {journal} {\bibinfo
   {journal} {ACS Nano}\ }\textbf {\bibinfo {volume} {7}},\ \bibinfo {pages}
  {7824} (\bibinfo {year} {2013}{\natexlab{a}})}\BibitemShut {NoStop}%
\bibitem [{\citenamefont {Kerker}\ \emph {et~al.}(1983)\citenamefont {Kerker},
  \citenamefont {Wang},\ and\ \citenamefont {Giles}}]{Kerker}%
  \BibitemOpen
  \bibfield  {author} {\bibinfo {author} {\bibfnamefont {M.}~\bibnamefont
  {Kerker}}, \bibinfo {author} {\bibfnamefont {D.-S.}\ \bibnamefont {Wang}}, \
  and\ \bibinfo {author} {\bibfnamefont {C.~L.}\ \bibnamefont {Giles}},\
  }\href@noop {} {\bibfield  {journal} {\bibinfo  {journal} {J. Opt. Soc. Am.}\
  }\textbf {\bibinfo {volume} {73}},\ \bibinfo {pages} {765} (\bibinfo {year}
  {1983})}\BibitemShut {NoStop}%
\bibitem [{\citenamefont {Zambrana-Puyalto}\ \emph
  {et~al.}(2013{\natexlab{a}})\citenamefont {Zambrana-Puyalto}, \citenamefont
  {Fern\'andez-Corbaton}, \citenamefont {Juan}, \citenamefont {Vidal},\ and\
  \citenamefont {Molina-Terriza}}]{ZambranaKerker}%
  \BibitemOpen
  \bibfield  {author} {\bibinfo {author} {\bibfnamefont {X.}~\bibnamefont
  {Zambrana-Puyalto}}, \bibinfo {author} {\bibfnamefont {I.}~\bibnamefont
  {Fern\'andez-Corbaton}}, \bibinfo {author} {\bibfnamefont {M.~L.}\
  \bibnamefont {Juan}}, \bibinfo {author} {\bibfnamefont {X.}~\bibnamefont
  {Vidal}}, \ and\ \bibinfo {author} {\bibfnamefont {G.}~\bibnamefont
  {Molina-Terriza}},\ }\href@noop {} {\bibfield  {journal} {\bibinfo  {journal}
  {Opt. Lett.}\ }\textbf {\bibinfo {volume} {38}},\ \bibinfo {pages} {1857}
  (\bibinfo {year} {2013}{\natexlab{a}})}\BibitemShut {NoStop}%
\bibitem [{\citenamefont {Nieto-Vesperinas}\ \emph
  {et~al.}(2011{\natexlab{a}})\citenamefont {Nieto-Vesperinas}, \citenamefont
  {G{\'o}mez-Medina},\ and\ \citenamefont {S{\'a}enz}}]{KerkerApp0}%
  \BibitemOpen
  \bibfield  {author} {\bibinfo {author} {\bibfnamefont {M.}~\bibnamefont
  {Nieto-Vesperinas}}, \bibinfo {author} {\bibfnamefont {R.}~\bibnamefont
  {G{\'o}mez-Medina}}, \ and\ \bibinfo {author} {\bibfnamefont {J.~J.}\
  \bibnamefont {S{\'a}enz}},\ }\href@noop {} {\bibfield  {journal} {\bibinfo
  {journal} {J. Opt. Soc. Am. A}\ }\textbf {\bibinfo {volume} {28}},\ \bibinfo
  {pages} {54} (\bibinfo {year} {2011}{\natexlab{a}})}\BibitemShut {NoStop}%
\bibitem [{\citenamefont {Garc\'{\i}a-Etxarri}\ and\ \citenamefont
  {Dionne}(2013)}]{KerkerApp1}%
  \BibitemOpen
  \bibfield  {author} {\bibinfo {author} {\bibfnamefont {A.}~\bibnamefont
  {Garc\'{\i}a-Etxarri}}\ and\ \bibinfo {author} {\bibfnamefont {J.~A.}\
  \bibnamefont {Dionne}},\ }\href@noop {} {\bibfield  {journal} {\bibinfo
  {journal} {Phys. Rev. B}\ }\textbf {\bibinfo {volume} {87}},\ \bibinfo
  {pages} {235409} (\bibinfo {year} {2013})}\BibitemShut {NoStop}%
\bibitem [{\citenamefont {Pfeiffer}\ and\ \citenamefont
  {Grbic}(2013)}]{KerkerApp2}%
  \BibitemOpen
  \bibfield  {author} {\bibinfo {author} {\bibfnamefont {C.}~\bibnamefont
  {Pfeiffer}}\ and\ \bibinfo {author} {\bibfnamefont {A.}~\bibnamefont
  {Grbic}},\ }\href@noop {} {\bibfield  {journal} {\bibinfo  {journal} {Phys.
  Rev. Lett.}\ }\textbf {\bibinfo {volume} {110}},\ \bibinfo {pages} {197401}
  (\bibinfo {year} {2013})}\BibitemShut {NoStop}%
\bibitem [{\citenamefont {Liu}\ and\ \citenamefont
  {Kivshar}(2018)}]{KerkerApp3}%
  \BibitemOpen
  \bibfield  {author} {\bibinfo {author} {\bibfnamefont {W.}~\bibnamefont
  {Liu}}\ and\ \bibinfo {author} {\bibfnamefont {Y.~S.}\ \bibnamefont
  {Kivshar}},\ }\href@noop {} {\bibfield  {journal} {\bibinfo  {journal} {Opt.
  Express}\ }\textbf {\bibinfo {volume} {26}},\ \bibinfo {pages} {13085}
  (\bibinfo {year} {2018})}\BibitemShut {NoStop}%
\bibitem [{\citenamefont {Solomon}\ \emph {et~al.}(2019)\citenamefont
  {Solomon}, \citenamefont {Hu}, \citenamefont {Lawrence}, \citenamefont
  {García-Etxarri},\ and\ \citenamefont {Dionne}}]{ACSSolomon}%
  \BibitemOpen
  \bibfield  {author} {\bibinfo {author} {\bibfnamefont {M.~L.}\ \bibnamefont
  {Solomon}}, \bibinfo {author} {\bibfnamefont {J.}~\bibnamefont {Hu}},
  \bibinfo {author} {\bibfnamefont {M.}~\bibnamefont {Lawrence}}, \bibinfo
  {author} {\bibfnamefont {A.}~\bibnamefont {García-Etxarri}}, \ and\ \bibinfo
  {author} {\bibfnamefont {J.~A.}\ \bibnamefont {Dionne}},\ }\href@noop {}
  {\bibfield  {journal} {\bibinfo  {journal} {ACS Photonics}\ }\textbf
  {\bibinfo {volume} {6}},\ \bibinfo {pages} {43} (\bibinfo {year}
  {2019})}\BibitemShut {NoStop}%
\bibitem [{\citenamefont {Lasa-Alonso}\ \emph
  {et~al.}(2020{\natexlab{a}})\citenamefont {Lasa-Alonso}, \citenamefont
  {Abujetas}, \citenamefont {Nodar}, \citenamefont {Dionne}, \citenamefont
  {S\'aenz}, \citenamefont {Molina-Terriza}, \citenamefont {Aizpurua},\ and\
  \citenamefont {Garc\'ia-Etxarri}}]{ACSLasa}%
  \BibitemOpen
  \bibfield  {author} {\bibinfo {author} {\bibfnamefont {J.}~\bibnamefont
  {Lasa-Alonso}}, \bibinfo {author} {\bibfnamefont {D.~R.}\ \bibnamefont
  {Abujetas}}, \bibinfo {author} {\bibfnamefont {A.}~\bibnamefont {Nodar}},
  \bibinfo {author} {\bibfnamefont {J.~A.}\ \bibnamefont {Dionne}}, \bibinfo
  {author} {\bibfnamefont {J.~J.}\ \bibnamefont {S\'aenz}}, \bibinfo {author}
  {\bibfnamefont {G.}~\bibnamefont {Molina-Terriza}}, \bibinfo {author}
  {\bibfnamefont {J.}~\bibnamefont {Aizpurua}}, \ and\ \bibinfo {author}
  {\bibfnamefont {A.}~\bibnamefont {Garc\'ia-Etxarri}},\ }\href@noop {}
  {\bibfield  {journal} {\bibinfo  {journal} {ACS Photonics}\ }\textbf
  {\bibinfo {volume} {7}},\ \bibinfo {pages} {2978–2986} (\bibinfo {year}
  {2020}{\natexlab{a}})}\BibitemShut {NoStop}%
\bibitem [{\citenamefont {Bliokh}\ \emph {et~al.}(2015)\citenamefont {Bliokh},
  \citenamefont {Rodr\'iguez-Fortuño}, \citenamefont {Nori},\ and\
  \citenamefont {Zayats}}]{Bliokh1}%
  \BibitemOpen
  \bibfield  {author} {\bibinfo {author} {\bibfnamefont {K.~Y.}\ \bibnamefont
  {Bliokh}}, \bibinfo {author} {\bibfnamefont {F.~J.}\ \bibnamefont
  {Rodr\'iguez-Fortuño}}, \bibinfo {author} {\bibfnamefont {F.}~\bibnamefont
  {Nori}}, \ and\ \bibinfo {author} {\bibfnamefont {A.~V.}\ \bibnamefont
  {Zayats}},\ }\href@noop {} {\bibfield  {journal} {\bibinfo  {journal} {Nature
  Photonics}\ }\textbf {\bibinfo {volume} {9}},\ \bibinfo {pages} {796}
  (\bibinfo {year} {2015})}\BibitemShut {NoStop}%
\bibitem [{\citenamefont {Allen}\ \emph {et~al.}(2003)\citenamefont {Allen},
  \citenamefont {Barnett},\ and\ \citenamefont {Padgett}}]{Barnett1}%
  \BibitemOpen
  \bibfield  {author} {\bibinfo {author} {\bibfnamefont {L.}~\bibnamefont
  {Allen}}, \bibinfo {author} {\bibfnamefont {S.~M.}\ \bibnamefont {Barnett}},
  \ and\ \bibinfo {author} {\bibfnamefont {M.~J.}\ \bibnamefont {Padgett}},\
  }\href@noop {} {\emph {\bibinfo {title} {Optical Angular Momentum}}}\
  (\bibinfo  {publisher} {Institute of Physics Publishing},\ \bibinfo {year}
  {2003})\BibitemShut {NoStop}%
\bibitem [{\citenamefont {Lasa-Alonso}\ \emph
  {et~al.}(2020{\natexlab{b}})\citenamefont {Lasa-Alonso}, \citenamefont
  {Molezuelas-Ferreras}, \citenamefont {Varga}, \citenamefont
  {Garc{\'{\i}}a-Etxarri}, \citenamefont {Giedke},\ and\ \citenamefont
  {Molina-Terriza}}]{SymProt}%
  \BibitemOpen
  \bibfield  {author} {\bibinfo {author} {\bibfnamefont {J.}~\bibnamefont
  {Lasa-Alonso}}, \bibinfo {author} {\bibfnamefont {M.}~\bibnamefont
  {Molezuelas-Ferreras}}, \bibinfo {author} {\bibfnamefont {J.~J.~M.}\
  \bibnamefont {Varga}}, \bibinfo {author} {\bibfnamefont {A.}~\bibnamefont
  {Garc{\'{\i}}a-Etxarri}}, \bibinfo {author} {\bibfnamefont {G.}~\bibnamefont
  {Giedke}}, \ and\ \bibinfo {author} {\bibfnamefont {G.}~\bibnamefont
  {Molina-Terriza}},\ }\href@noop {} {\bibfield  {journal} {\bibinfo  {journal}
  {New Journal of Physics}\ }\textbf {\bibinfo {volume} {22}},\ \bibinfo
  {pages} {123010} (\bibinfo {year} {2020}{\natexlab{b}})}\BibitemShut
  {NoStop}%
\bibitem [{\citenamefont {Mehta}\ \emph {et~al.}(2006)\citenamefont {Mehta},
  \citenamefont {Patel}, \citenamefont {Desai}, \citenamefont {Upadhyay},\ and\
  \citenamefont {Parekh}}]{Optimal3}%
  \BibitemOpen
  \bibfield  {author} {\bibinfo {author} {\bibfnamefont {R.~V.}\ \bibnamefont
  {Mehta}}, \bibinfo {author} {\bibfnamefont {R.}~\bibnamefont {Patel}},
  \bibinfo {author} {\bibfnamefont {R.}~\bibnamefont {Desai}}, \bibinfo
  {author} {\bibfnamefont {R.~V.}\ \bibnamefont {Upadhyay}}, \ and\ \bibinfo
  {author} {\bibfnamefont {K.}~\bibnamefont {Parekh}},\ }\href@noop {}
  {\bibfield  {journal} {\bibinfo  {journal} {Phys. Rev. Lett.}\ }\textbf
  {\bibinfo {volume} {96}},\ \bibinfo {pages} {127402} (\bibinfo {year}
  {2006})}\BibitemShut {NoStop}%
\bibitem [{\citenamefont {Garc\'{\i}a-C\'amara}\ \emph
  {et~al.}(2007)\citenamefont {Garc\'{\i}a-C\'amara}, \citenamefont {Moreno},
  \citenamefont {Gonz\'alez},\ and\ \citenamefont {Saiz}}]{Optimal31}%
  \BibitemOpen
  \bibfield  {author} {\bibinfo {author} {\bibfnamefont {B.}~\bibnamefont
  {Garc\'{\i}a-C\'amara}}, \bibinfo {author} {\bibfnamefont {F.}~\bibnamefont
  {Moreno}}, \bibinfo {author} {\bibfnamefont {F.}~\bibnamefont {Gonz\'alez}},
  \ and\ \bibinfo {author} {\bibfnamefont {J.~M.}\ \bibnamefont {Saiz}},\
  }\href@noop {} {\bibfield  {journal} {\bibinfo  {journal} {Phys. Rev. Lett.}\
  }\textbf {\bibinfo {volume} {98}},\ \bibinfo {pages} {179701} (\bibinfo
  {year} {2007})}\BibitemShut {NoStop}%
\bibitem [{\citenamefont {Ramachandran}\ and\ \citenamefont
  {Kumar}(2008)}]{Optimal32}%
  \BibitemOpen
  \bibfield  {author} {\bibinfo {author} {\bibfnamefont {H.}~\bibnamefont
  {Ramachandran}}\ and\ \bibinfo {author} {\bibfnamefont {N.}~\bibnamefont
  {Kumar}},\ }\href@noop {} {\bibfield  {journal} {\bibinfo  {journal} {Phys.
  Rev. Lett.}\ }\textbf {\bibinfo {volume} {100}},\ \bibinfo {pages} {229703}
  (\bibinfo {year} {2008})}\BibitemShut {NoStop}%
\bibitem [{\citenamefont {Garc{\'\i}a-C{\'a}mara}\ \emph
  {et~al.}(2008)\citenamefont {Garc{\'\i}a-C{\'a}mara}, \citenamefont
  {Gonz{\'a}lez}, \citenamefont {Moreno},\ and\ \citenamefont
  {Saiz}}]{CamaraExcep}%
  \BibitemOpen
  \bibfield  {author} {\bibinfo {author} {\bibfnamefont {B.}~\bibnamefont
  {Garc{\'\i}a-C{\'a}mara}}, \bibinfo {author} {\bibfnamefont {F.}~\bibnamefont
  {Gonz{\'a}lez}}, \bibinfo {author} {\bibfnamefont {F.}~\bibnamefont
  {Moreno}}, \ and\ \bibinfo {author} {\bibfnamefont {J.~M.}\ \bibnamefont
  {Saiz}},\ }\href@noop {} {\bibfield  {journal} {\bibinfo  {journal} {JOSA A}\
  }\textbf {\bibinfo {volume} {25}},\ \bibinfo {pages} {2875} (\bibinfo {year}
  {2008})}\BibitemShut {NoStop}%
\bibitem [{\citenamefont {Al\'u}\ and\ \citenamefont
  {Engheta}(2010)}]{AluEngheta}%
  \BibitemOpen
  \bibfield  {author} {\bibinfo {author} {\bibfnamefont {A.}~\bibnamefont
  {Al\'u}}\ and\ \bibinfo {author} {\bibfnamefont {N.}~\bibnamefont
  {Engheta}},\ }\href@noop {} {\bibfield  {journal} {\bibinfo  {journal}
  {Journal of Nanophotonics}\ }\textbf {\bibinfo {volume} {4}},\ \bibinfo
  {pages} {041590} (\bibinfo {year} {2010})}\BibitemShut {NoStop}%
\bibitem [{\citenamefont {Garc\'{i}a-C\'{a}mara}\ \emph
  {et~al.}(2011)\citenamefont {Garc\'{i}a-C\'{a}mara}, \citenamefont {de~la
  Osa}, \citenamefont {Saiz}, \citenamefont {Gonz\'{a}lez},\ and\ \citenamefont
  {Moreno}}]{Optimal4}%
  \BibitemOpen
  \bibfield  {author} {\bibinfo {author} {\bibfnamefont {B.}~\bibnamefont
  {Garc\'{i}a-C\'{a}mara}}, \bibinfo {author} {\bibfnamefont {R.~A.}\
  \bibnamefont {de~la Osa}}, \bibinfo {author} {\bibfnamefont {J.~M.}\
  \bibnamefont {Saiz}}, \bibinfo {author} {\bibfnamefont {F.}~\bibnamefont
  {Gonz\'{a}lez}}, \ and\ \bibinfo {author} {\bibfnamefont {F.}~\bibnamefont
  {Moreno}},\ }\href@noop {} {\bibfield  {journal} {\bibinfo  {journal} {Opt.
  Lett.}\ }\textbf {\bibinfo {volume} {36}},\ \bibinfo {pages} {728} (\bibinfo
  {year} {2011})}\BibitemShut {NoStop}%
\bibitem [{\citenamefont {Geffrin}\ \emph {et~al.}(2012)\citenamefont
  {Geffrin}, \citenamefont {Garc{\'i}a-C{\'a}mara}, \citenamefont
  {G{\'o}mez-Medina}, \citenamefont {Albella}, \citenamefont
  {Froufe-P{\'e}rez}, \citenamefont {Eyraud}, \citenamefont {Litman},
  \citenamefont {Vaillon}, \citenamefont {Gonz{\'a}lez}, \citenamefont
  {Nieto-Vesperinas}, \citenamefont {S{\'a}enz},\ and\ \citenamefont
  {Moreno}}]{KerkerRef0}%
  \BibitemOpen
  \bibfield  {author} {\bibinfo {author} {\bibfnamefont {J.}~\bibnamefont
  {Geffrin}}, \bibinfo {author} {\bibfnamefont {B.}~\bibnamefont
  {Garc{\'i}a-C{\'a}mara}}, \bibinfo {author} {\bibfnamefont {R.}~\bibnamefont
  {G{\'o}mez-Medina}}, \bibinfo {author} {\bibfnamefont {P.}~\bibnamefont
  {Albella}}, \bibinfo {author} {\bibfnamefont {L.~S.}\ \bibnamefont
  {Froufe-P{\'e}rez}}, \bibinfo {author} {\bibfnamefont {C.}~\bibnamefont
  {Eyraud}}, \bibinfo {author} {\bibfnamefont {A.}~\bibnamefont {Litman}},
  \bibinfo {author} {\bibfnamefont {R.}~\bibnamefont {Vaillon}}, \bibinfo
  {author} {\bibfnamefont {F.}~\bibnamefont {Gonz{\'a}lez}}, \bibinfo {author}
  {\bibfnamefont {M.}~\bibnamefont {Nieto-Vesperinas}}, \bibinfo {author}
  {\bibfnamefont {J.~J.}\ \bibnamefont {S{\'a}enz}}, \ and\ \bibinfo {author}
  {\bibfnamefont {F.}~\bibnamefont {Moreno}},\ }\href@noop {} {\bibfield
  {journal} {\bibinfo  {journal} {Nature Communications}\ }\textbf {\bibinfo
  {volume} {3}} (\bibinfo {year} {2012})}\BibitemShut {NoStop}%
\bibitem [{\citenamefont {Lee}\ \emph {et~al.}(2018)\citenamefont {Lee},
  \citenamefont {Miroshnichenko},\ and\ \citenamefont {Lee}}]{MiroZero}%
  \BibitemOpen
  \bibfield  {author} {\bibinfo {author} {\bibfnamefont {J.~Y.}\ \bibnamefont
  {Lee}}, \bibinfo {author} {\bibfnamefont {A.~E.}\ \bibnamefont
  {Miroshnichenko}}, \ and\ \bibinfo {author} {\bibfnamefont {R.-K.}\
  \bibnamefont {Lee}},\ }\href@noop {} {\bibfield  {journal} {\bibinfo
  {journal} {Optics express}\ }\textbf {\bibinfo {volume} {26}},\ \bibinfo
  {pages} {30393} (\bibinfo {year} {2018})}\BibitemShut {NoStop}%
\bibitem [{\citenamefont {Zambrana-Puyalto}\ \emph
  {et~al.}(2013{\natexlab{b}})\citenamefont {Zambrana-Puyalto}, \citenamefont
  {Vidal}, \citenamefont {Juan},\ and\ \citenamefont
  {Molina-Terriza}}]{DualAntidualModes}%
  \BibitemOpen
  \bibfield  {author} {\bibinfo {author} {\bibfnamefont {X.}~\bibnamefont
  {Zambrana-Puyalto}}, \bibinfo {author} {\bibfnamefont {X.}~\bibnamefont
  {Vidal}}, \bibinfo {author} {\bibfnamefont {M.~L.}\ \bibnamefont {Juan}}, \
  and\ \bibinfo {author} {\bibfnamefont {G.}~\bibnamefont {Molina-Terriza}},\
  }\href@noop {} {\bibfield  {journal} {\bibinfo  {journal} {Optics Express}\
  }\textbf {\bibinfo {volume} {21}},\ \bibinfo {pages} {17520} (\bibinfo {year}
  {2013}{\natexlab{b}})}\BibitemShut {NoStop}%
\bibitem [{\citenamefont {Shen}\ \emph {et~al.}(2017)\citenamefont {Shen},
  \citenamefont {An}, \citenamefont {Tao}, \citenamefont {Zhou}, \citenamefont
  {Jiang},\ and\ \citenamefont {Guo}}]{ShenGain}%
  \BibitemOpen
  \bibfield  {author} {\bibinfo {author} {\bibfnamefont {F.}~\bibnamefont
  {Shen}}, \bibinfo {author} {\bibfnamefont {N.}~\bibnamefont {An}}, \bibinfo
  {author} {\bibfnamefont {Y.}~\bibnamefont {Tao}}, \bibinfo {author}
  {\bibfnamefont {H.}~\bibnamefont {Zhou}}, \bibinfo {author} {\bibfnamefont
  {Z.}~\bibnamefont {Jiang}}, \ and\ \bibinfo {author} {\bibfnamefont
  {Z.}~\bibnamefont {Guo}},\ }\href@noop {} {\bibfield  {journal} {\bibinfo
  {journal} {Nanophotonics}\ }\textbf {\bibinfo {volume} {6}},\ \bibinfo
  {pages} {1063} (\bibinfo {year} {2017})}\BibitemShut {NoStop}%
\bibitem [{\citenamefont {Sanz-Fern{\'a}ndez}\ \emph
  {et~al.}(2021)\citenamefont {Sanz-Fern{\'a}ndez}, \citenamefont
  {Molezuelas-Ferreras}, \citenamefont {Lasa-Alonso}, \citenamefont {de~Sousa},
  \citenamefont {Zambrana-Puyalto},\ and\ \citenamefont
  {Olmos-Trigo}}]{CrisAnapole}%
  \BibitemOpen
  \bibfield  {author} {\bibinfo {author} {\bibfnamefont {C.}~\bibnamefont
  {Sanz-Fern{\'a}ndez}}, \bibinfo {author} {\bibfnamefont {M.}~\bibnamefont
  {Molezuelas-Ferreras}}, \bibinfo {author} {\bibfnamefont {J.}~\bibnamefont
  {Lasa-Alonso}}, \bibinfo {author} {\bibfnamefont {N.}~\bibnamefont
  {de~Sousa}}, \bibinfo {author} {\bibfnamefont {X.}~\bibnamefont
  {Zambrana-Puyalto}}, \ and\ \bibinfo {author} {\bibfnamefont
  {J.}~\bibnamefont {Olmos-Trigo}},\ }\href@noop {} {\bibfield  {journal}
  {\bibinfo  {journal} {Laser \& Photonics Reviews}\ }\textbf {\bibinfo
  {volume} {15}},\ \bibinfo {pages} {2100035} (\bibinfo {year}
  {2021})}\BibitemShut {NoStop}%
\bibitem [{\citenamefont {Ali}(2022)}]{RAliGain}%
  \BibitemOpen
  \bibfield  {author} {\bibinfo {author} {\bibfnamefont {R.}~\bibnamefont
  {Ali}},\ }\href@noop {} {\bibfield  {journal} {\bibinfo  {journal} {Physical
  Review A}\ }\textbf {\bibinfo {volume} {106}},\ \bibinfo {pages} {063508}
  (\bibinfo {year} {2022})}\BibitemShut {NoStop}%
\bibitem [{\citenamefont {Bohren}\ and\ \citenamefont
  {Huffman}(1998)}]{Bohren}%
  \BibitemOpen
  \bibfield  {author} {\bibinfo {author} {\bibfnamefont {C.~F.}\ \bibnamefont
  {Bohren}}\ and\ \bibinfo {author} {\bibfnamefont {D.~R.}\ \bibnamefont
  {Huffman}},\ }\href@noop {} {\emph {\bibinfo {title} {Absorption and
  scattering of light by small particles}}}\ (\bibinfo  {publisher} {John Wiley
  and Sons, Inc.},\ \bibinfo {year} {1998})\BibitemShut {NoStop}%
\bibitem [{\citenamefont {Novotny}\ and\ \citenamefont
  {Hetch}(2006)}]{Novotny}%
  \BibitemOpen
  \bibfield  {author} {\bibinfo {author} {\bibfnamefont {L.}~\bibnamefont
  {Novotny}}\ and\ \bibinfo {author} {\bibfnamefont {B.}~\bibnamefont
  {Hetch}},\ }\href@noop {} {\emph {\bibinfo {title} {Principles of
  Nano-optics}}}\ (\bibinfo  {publisher} {Cambridge University Press},\
  \bibinfo {year} {2006})\BibitemShut {NoStop}%
\bibitem [{\citenamefont {Jackson}(1999)}]{Jackson}%
  \BibitemOpen
  \bibfield  {author} {\bibinfo {author} {\bibfnamefont {J.~D.}\ \bibnamefont
  {Jackson}},\ }\href@noop {} {\emph {\bibinfo {title} {Classical
  Electrodynamics}}}\ (\bibinfo  {publisher} {John Wiley and Sons, Inc.},\
  \bibinfo {year} {1999})\BibitemShut {NoStop}%
\bibitem [{\citenamefont {Fern\'andez-Corbaton}(2013)}]{ForBackCorbaton}%
  \BibitemOpen
  \bibfield  {author} {\bibinfo {author} {\bibfnamefont {I.}~\bibnamefont
  {Fern\'andez-Corbaton}},\ }\href@noop {} {\bibfield  {journal} {\bibinfo
  {journal} {Optics Express}\ }\textbf {\bibinfo {volume} {21}},\ \bibinfo
  {pages} {29885} (\bibinfo {year} {2013})}\BibitemShut {NoStop}%
\bibitem [{\citenamefont {Weber}(1901)}]{WeberRiemann}%
  \BibitemOpen
  \bibfield  {author} {\bibinfo {author} {\bibfnamefont {H.}~\bibnamefont
  {Weber}},\ }\href@noop {} {\emph {\bibinfo {title} {Die Partiellen
  Differential-Gleichungen der Mathematischen Physik}}}\ (\bibinfo  {publisher}
  {Braunschweig},\ \bibinfo {year} {1901})\BibitemShut {NoStop}%
\bibitem [{\citenamefont {Silberstein}(1907)}]{Silberstein}%
  \BibitemOpen
  \bibfield  {author} {\bibinfo {author} {\bibfnamefont {L.}~\bibnamefont
  {Silberstein}},\ }\href@noop {} {\bibfield  {journal} {\bibinfo  {journal}
  {Annalen der Physik}\ }\textbf {\bibinfo {volume} {327}},\ \bibinfo {pages}
  {579} (\bibinfo {year} {1907})}\BibitemShut {NoStop}%
\bibitem [{\citenamefont {Bialynicki-Birula}\ and\ \citenamefont
  {Bialynicka-Birula}(2013)}]{RoleRS}%
  \BibitemOpen
  \bibfield  {author} {\bibinfo {author} {\bibfnamefont {I.}~\bibnamefont
  {Bialynicki-Birula}}\ and\ \bibinfo {author} {\bibfnamefont {Z.}~\bibnamefont
  {Bialynicka-Birula}},\ }\href@noop {} {\bibfield  {journal} {\bibinfo
  {journal} {Journal of Physics A: Mathematical and Theoretical}\ }\textbf
  {\bibinfo {volume} {46}},\ \bibinfo {pages} {053001} (\bibinfo {year}
  {2013})}\BibitemShut {NoStop}%
\bibitem [{\citenamefont {Barnett}(2014)}]{BarnettPhotonHam}%
  \BibitemOpen
  \bibfield  {author} {\bibinfo {author} {\bibfnamefont {S.~M.}\ \bibnamefont
  {Barnett}},\ }\href {\doibase 10.1088/1367-2630/16/9/093008} {\bibfield
  {journal} {\bibinfo  {journal} {New Journal of Physics}\ }\textbf {\bibinfo
  {volume} {16}},\ \bibinfo {pages} {093008} (\bibinfo {year}
  {2014})}\BibitemShut {NoStop}%
\bibitem [{\citenamefont {Bialynicki-Birula}\ and\ \citenamefont
  {Bialynicka-Birula}(2017)}]{IZBirula}%
  \BibitemOpen
  \bibfield  {author} {\bibinfo {author} {\bibfnamefont {I.}~\bibnamefont
  {Bialynicki-Birula}}\ and\ \bibinfo {author} {\bibfnamefont {Z.}~\bibnamefont
  {Bialynicka-Birula}},\ }\href@noop {} {\bibfield  {journal} {\bibinfo
  {journal} {Journal of Optics}\ }\textbf {\bibinfo {volume} {19}},\ \bibinfo
  {pages} {125201} (\bibinfo {year} {2017})}\BibitemShut {NoStop}%
\bibitem [{\citenamefont {Kiessling}\ and\ \citenamefont
  {Tahvildar-Zadeh}(2018)}]{PhotonHam1}%
  \BibitemOpen
  \bibfield  {author} {\bibinfo {author} {\bibfnamefont {M.~K.-H.}\
  \bibnamefont {Kiessling}}\ and\ \bibinfo {author} {\bibfnamefont {A.~S.}\
  \bibnamefont {Tahvildar-Zadeh}},\ }\href@noop {} {\bibfield  {journal}
  {\bibinfo  {journal} {Journal of Mathematical Physics}\ }\textbf {\bibinfo
  {volume} {59}},\ \bibinfo {pages} {112302} (\bibinfo {year}
  {2018})}\BibitemShut {NoStop}%
\bibitem [{\citenamefont {Babicheva}\ and\ \citenamefont
  {Evlyukhin}(2017)}]{LatticeKerker}%
  \BibitemOpen
  \bibfield  {author} {\bibinfo {author} {\bibfnamefont {V.~E.}\ \bibnamefont
  {Babicheva}}\ and\ \bibinfo {author} {\bibfnamefont {A.~B.}\ \bibnamefont
  {Evlyukhin}},\ }\href@noop {} {\bibfield  {journal} {\bibinfo  {journal}
  {Laser \& Photonics Reviews}\ }\textbf {\bibinfo {volume} {11}},\ \bibinfo
  {pages} {1700132} (\bibinfo {year} {2017})}\BibitemShut {NoStop}%
\bibitem [{\citenamefont {Staude}\ \emph
  {et~al.}(2013{\natexlab{b}})\citenamefont {Staude}, \citenamefont
  {Miroshnichenko}, \citenamefont {Decker}, \citenamefont {Fofang},
  \citenamefont {Liu}, \citenamefont {Gonzales}, \citenamefont {Dominguez},
  \citenamefont {Luk}, \citenamefont {Neshev}, \citenamefont {Brener},\ and\
  \citenamefont {Kivshar}}]{KivsharHuygens2}%
  \BibitemOpen
  \bibfield  {author} {\bibinfo {author} {\bibfnamefont {I.}~\bibnamefont
  {Staude}}, \bibinfo {author} {\bibfnamefont {A.~E.}\ \bibnamefont
  {Miroshnichenko}}, \bibinfo {author} {\bibfnamefont {M.}~\bibnamefont
  {Decker}}, \bibinfo {author} {\bibfnamefont {N.~T.}\ \bibnamefont {Fofang}},
  \bibinfo {author} {\bibfnamefont {S.}~\bibnamefont {Liu}}, \bibinfo {author}
  {\bibfnamefont {E.}~\bibnamefont {Gonzales}}, \bibinfo {author}
  {\bibfnamefont {J.}~\bibnamefont {Dominguez}}, \bibinfo {author}
  {\bibfnamefont {T.~S.}\ \bibnamefont {Luk}}, \bibinfo {author} {\bibfnamefont
  {D.~N.}\ \bibnamefont {Neshev}}, \bibinfo {author} {\bibfnamefont
  {I.}~\bibnamefont {Brener}}, \ and\ \bibinfo {author} {\bibfnamefont
  {Y.}~\bibnamefont {Kivshar}},\ }\href@noop {} {\bibfield  {journal} {\bibinfo
   {journal} {ACS Nano}\ }\textbf {\bibinfo {volume} {7}},\ \bibinfo {pages}
  {7824} (\bibinfo {year} {2013}{\natexlab{b}})}\BibitemShut {NoStop}%
\bibitem [{\citenamefont {Pfeiffer}\ \emph {et~al.}(2014)\citenamefont
  {Pfeiffer}, \citenamefont {Emani}, \citenamefont {Shaltout}, \citenamefont
  {Boltasseva}, \citenamefont {Shalaev},\ and\ \citenamefont
  {Grbic}}]{PfeifferGrbic2}%
  \BibitemOpen
  \bibfield  {author} {\bibinfo {author} {\bibfnamefont {C.}~\bibnamefont
  {Pfeiffer}}, \bibinfo {author} {\bibfnamefont {N.~K.}\ \bibnamefont {Emani}},
  \bibinfo {author} {\bibfnamefont {A.~M.}\ \bibnamefont {Shaltout}}, \bibinfo
  {author} {\bibfnamefont {A.}~\bibnamefont {Boltasseva}}, \bibinfo {author}
  {\bibfnamefont {V.~M.}\ \bibnamefont {Shalaev}}, \ and\ \bibinfo {author}
  {\bibfnamefont {A.}~\bibnamefont {Grbic}},\ }\href@noop {} {\bibfield
  {journal} {\bibinfo  {journal} {Nano Letters}\ }\textbf {\bibinfo {volume}
  {14}},\ \bibinfo {pages} {2491} (\bibinfo {year} {2014})}\BibitemShut
  {NoStop}%
\bibitem [{\citenamefont {Decker}\ \emph {et~al.}(2015)\citenamefont {Decker},
  \citenamefont {Staude}, \citenamefont {Falkner}, \citenamefont {Dominguez},
  \citenamefont {Neshev}, \citenamefont {Brener}, \citenamefont {Pertsch},\
  and\ \citenamefont {Kivshar}}]{KivsharHuygens}%
  \BibitemOpen
  \bibfield  {author} {\bibinfo {author} {\bibfnamefont {M.}~\bibnamefont
  {Decker}}, \bibinfo {author} {\bibfnamefont {I.}~\bibnamefont {Staude}},
  \bibinfo {author} {\bibfnamefont {M.}~\bibnamefont {Falkner}}, \bibinfo
  {author} {\bibfnamefont {J.}~\bibnamefont {Dominguez}}, \bibinfo {author}
  {\bibfnamefont {D.~N.}\ \bibnamefont {Neshev}}, \bibinfo {author}
  {\bibfnamefont {I.}~\bibnamefont {Brener}}, \bibinfo {author} {\bibfnamefont
  {T.}~\bibnamefont {Pertsch}}, \ and\ \bibinfo {author} {\bibfnamefont
  {Y.~S.}\ \bibnamefont {Kivshar}},\ }\href@noop {} {\bibfield  {journal}
  {\bibinfo  {journal} {Advanced Optical Materials}\ }\textbf {\bibinfo
  {volume} {3}},\ \bibinfo {pages} {813} (\bibinfo {year} {2015})}\BibitemShut
  {NoStop}%
\bibitem [{\citenamefont {Giles}\ and\ \citenamefont {Wild}(1982)}]{GilesWild}%
  \BibitemOpen
  \bibfield  {author} {\bibinfo {author} {\bibfnamefont {C.~L.}\ \bibnamefont
  {Giles}}\ and\ \bibinfo {author} {\bibfnamefont {W.~J.}\ \bibnamefont
  {Wild}},\ }\href@noop {} {\bibfield  {journal} {\bibinfo  {journal} {Applied
  Physics Letters}\ }\textbf {\bibinfo {volume} {40}},\ \bibinfo {pages} {210}
  (\bibinfo {year} {1982})}\BibitemShut {NoStop}%
\bibitem [{\citenamefont {Lakhtakia}(1990)}]{Lakhtakia}%
  \BibitemOpen
  \bibfield  {author} {\bibinfo {author} {\bibfnamefont {A.}~\bibnamefont
  {Lakhtakia}},\ }\href@noop {} {\bibfield  {journal} {\bibinfo  {journal}
  {International Journal of Infrared and Milimiter Waves}\ }\textbf {\bibinfo
  {volume} {11}},\ \bibinfo {pages} {1407} (\bibinfo {year}
  {1990})}\BibitemShut {NoStop}%
\bibitem [{\citenamefont {Lasa-Alonso}\ \emph {et~al.}(2022)\citenamefont
  {Lasa-Alonso}, \citenamefont {Olmos-Trigo}, \citenamefont {García-Etxarri},\
  and\ \citenamefont {Molina-Terriza}}]{Correlations}%
  \BibitemOpen
  \bibfield  {author} {\bibinfo {author} {\bibfnamefont {J.}~\bibnamefont
  {Lasa-Alonso}}, \bibinfo {author} {\bibfnamefont {J.}~\bibnamefont
  {Olmos-Trigo}}, \bibinfo {author} {\bibfnamefont {A.}~\bibnamefont
  {García-Etxarri}}, \ and\ \bibinfo {author} {\bibfnamefont {G.}~\bibnamefont
  {Molina-Terriza}},\ }\href@noop {} {\bibfield  {journal} {\bibinfo  {journal}
  {Mater. Adv.}\ }\textbf {\bibinfo {volume} {3}},\ \bibinfo {pages} {4179}
  (\bibinfo {year} {2022})}\BibitemShut {NoStop}%
\bibitem [{\citenamefont {Olmos-Trigo}\ \emph
  {et~al.}(2020{\natexlab{a}})\citenamefont {Olmos-Trigo}, \citenamefont
  {Sanz-Fern\'andez}, \citenamefont {Abujetas}, \citenamefont {Lasa-Alonso},
  \citenamefont {de~Sousa}, \citenamefont {Garc\'ia-Etxarri}, \citenamefont
  {S\'anchez-Gil}, \citenamefont {Molina-Terriza},\ and\ \citenamefont
  {S\'aenz}}]{PRLJorge}%
  \BibitemOpen
  \bibfield  {author} {\bibinfo {author} {\bibfnamefont {J.}~\bibnamefont
  {Olmos-Trigo}}, \bibinfo {author} {\bibfnamefont {C.}~\bibnamefont
  {Sanz-Fern\'andez}}, \bibinfo {author} {\bibfnamefont {D.~R.}\ \bibnamefont
  {Abujetas}}, \bibinfo {author} {\bibfnamefont {J.}~\bibnamefont
  {Lasa-Alonso}}, \bibinfo {author} {\bibfnamefont {N.}~\bibnamefont
  {de~Sousa}}, \bibinfo {author} {\bibfnamefont {A.}~\bibnamefont
  {Garc\'ia-Etxarri}}, \bibinfo {author} {\bibfnamefont {J.~A.}\ \bibnamefont
  {S\'anchez-Gil}}, \bibinfo {author} {\bibfnamefont {G.}~\bibnamefont
  {Molina-Terriza}}, \ and\ \bibinfo {author} {\bibfnamefont {J.~J.}\
  \bibnamefont {S\'aenz}},\ }\href@noop {} {\bibfield  {journal} {\bibinfo
  {journal} {Phys. Rev. Lett.}\ }\textbf {\bibinfo {volume} {125}},\ \bibinfo
  {pages} {073205} (\bibinfo {year} {2020}{\natexlab{a}})}\BibitemShut
  {NoStop}%
\bibitem [{\citenamefont {Cohen-Tannoudji}\ \emph {et~al.}(1977)\citenamefont
  {Cohen-Tannoudji}, \citenamefont {Diu},\ and\ \citenamefont
  {Laloë}}]{CohenTannoudji}%
  \BibitemOpen
  \bibfield  {author} {\bibinfo {author} {\bibfnamefont {C.}~\bibnamefont
  {Cohen-Tannoudji}}, \bibinfo {author} {\bibfnamefont {B.}~\bibnamefont
  {Diu}}, \ and\ \bibinfo {author} {\bibfnamefont {F.}~\bibnamefont {Laloë}},\
  }\href@noop {} {\emph {\bibinfo {title} {Quantum Mechanics}}}\ (\bibinfo
  {publisher} {John Wiley and Sons, Inc.},\ \bibinfo {year} {1977})\BibitemShut
  {NoStop}%
\bibitem [{\citenamefont {Bahcall}\ \emph {et~al.}(2018)\citenamefont
  {Bahcall}, \citenamefont {Davis}, \citenamefont {Parker}, \citenamefont
  {Smirnov},\ and\ \citenamefont {Ulrich}}]{MSWSolar}%
  \BibitemOpen
  \bibinfo {editor} {\bibfnamefont {J.~N.}\ \bibnamefont {Bahcall}}, \bibinfo
  {editor} {\bibfnamefont {R.}~\bibnamefont {Davis}}, \bibinfo {editor}
  {\bibfnamefont {P.}~\bibnamefont {Parker}}, \bibinfo {editor} {\bibfnamefont
  {A.}~\bibnamefont {Smirnov}}, \ and\ \bibinfo {editor} {\bibfnamefont
  {R.}~\bibnamefont {Ulrich}},\ eds.,\ \href@noop {} {\emph {\bibinfo {title}
  {Solar Neutrinos: the First Thirty Years}}}\ (\bibinfo  {publisher} {CRC
  Press},\ \bibinfo {year} {2018})\BibitemShut {NoStop}%
\bibitem [{\citenamefont {Nieto-Vesperinas}\ \emph
  {et~al.}(2011{\natexlab{b}})\citenamefont {Nieto-Vesperinas}, \citenamefont
  {Gomez-Medina},\ and\ \citenamefont {Saenz}}]{NietoVesperinasK2}%
  \BibitemOpen
  \bibfield  {author} {\bibinfo {author} {\bibfnamefont {M.}~\bibnamefont
  {Nieto-Vesperinas}}, \bibinfo {author} {\bibfnamefont {R.}~\bibnamefont
  {Gomez-Medina}}, \ and\ \bibinfo {author} {\bibfnamefont {J.~J.}\
  \bibnamefont {Saenz}},\ }\href@noop {} {\bibfield  {journal} {\bibinfo
  {journal} {J. Opt. Soc. Am. A}\ }\textbf {\bibinfo {volume} {28}},\ \bibinfo
  {pages} {54} (\bibinfo {year} {2011}{\natexlab{b}})}\BibitemShut {NoStop}%
\bibitem [{\citenamefont {Lee}\ \emph {et~al.}(2017)\citenamefont {Lee},
  \citenamefont {Miroshnichenko},\ and\ \citenamefont {Lee}}]{MiroshisenkoK2}%
  \BibitemOpen
  \bibfield  {author} {\bibinfo {author} {\bibfnamefont {J.~Y.}\ \bibnamefont
  {Lee}}, \bibinfo {author} {\bibfnamefont {A.~E.}\ \bibnamefont
  {Miroshnichenko}}, \ and\ \bibinfo {author} {\bibfnamefont {R.-K.}\
  \bibnamefont {Lee}},\ }\href@noop {} {\bibfield  {journal} {\bibinfo
  {journal} {Phys. Rev. A}\ }\textbf {\bibinfo {volume} {96}},\ \bibinfo
  {pages} {043846} (\bibinfo {year} {2017})}\BibitemShut {NoStop}%
\bibitem [{\citenamefont {Olmos-Trigo}\ \emph
  {et~al.}(2020{\natexlab{b}})\citenamefont {Olmos-Trigo}, \citenamefont
  {Abujetas}, \citenamefont {Sanz-Fern\'andez}, \citenamefont {S\'anchez-Gil},\
  and\ \citenamefont {S\'aenz}}]{PRROlmos}%
  \BibitemOpen
  \bibfield  {author} {\bibinfo {author} {\bibfnamefont {J.}~\bibnamefont
  {Olmos-Trigo}}, \bibinfo {author} {\bibfnamefont {D.~R.}\ \bibnamefont
  {Abujetas}}, \bibinfo {author} {\bibfnamefont {C.}~\bibnamefont
  {Sanz-Fern\'andez}}, \bibinfo {author} {\bibfnamefont {J.~A.}\ \bibnamefont
  {S\'anchez-Gil}}, \ and\ \bibinfo {author} {\bibfnamefont {J.~J.}\
  \bibnamefont {S\'aenz}},\ }\href@noop {} {\bibfield  {journal} {\bibinfo
  {journal} {Physical Review Research}\ }\textbf {\bibinfo {volume} {2}},\
  \bibinfo {pages} {013225} (\bibinfo {year} {2020}{\natexlab{b}})}\BibitemShut
  {NoStop}%
\end{thebibliography}%

\end{document}